\documentclass[fleqn,usenatbib,useAMS]{mnras}

\usepackage{graphicx}	
\usepackage{amsmath}	
\usepackage{amssymb}	
\usepackage{multicol}        
\usepackage{bm}		
\usepackage{float}
\usepackage{gensymb}
\usepackage[usenames, dvipsnames]{color}
\usepackage{subfigure}
\usepackage{commath}
\usepackage{enumitem}
\setlist[enumerate]{leftmargin=\parindent}
\setlist[itemize]{leftmargin=\parindent}
\usepackage{longtable}

\newcommand{\NGTS}{NGTS}
\newcommand{\SPECULOOS}{SPECULOOS}
\newcommand{\HARPS}{HARPS}

\newcommand{\TESS}{TESS}
\newcommand{\Gaia}{\textit{Gaia}}

\newcommand{\NGTSID}{\mbox{NGTS~J061746.7-354222.9}}
\newcommand{\TWOMASSID}{\mbox{2MASS~J06174675-3542230}}
\newcommand{\GaiaID}{\mbox{\textit{Gaia}~2885350546895266432} (DR2)} 
\newcommand{\NGTSIDnr}{\mbox{J061746.7-354222.9}}
\newcommand{\TWOMASSIDnr}{\mbox{J06174675-3542230}}
\newcommand{\GaiaIDnr}{\mbox{2885350546895266432}} 
\newcommand{\Nplanet}{\mbox{NGTS-3Ab}}
\newcommand{\Ntarget}{\mbox{NGTS-3}}
\newcommand{\NstarA}{\mbox{NGTS-3A}}
\newcommand{\NstarB}{\mbox{NGTS-3B}}
\newcommand{\Nperiod}{1.675}

\newcommand{\RVsysA}{$RV_\mathrm{sys,A}$$=8.566\pm0.049$} 
\newcommand{\RVsysB}{$RV_\mathrm{sys,B}$$=9.032_{-0.064}^{+0.085}$} 
\newcommand{\RVK}{$K$$=-0.404\pm0.035$}

\newcommand{\TeffB}{$T_\mathrm{eff,B}$$=5230_{-220}^{+190}$} 
\newcommand{\RB}{$R_\mathrm{B}$$=0.77_{-0.16}^{+0.22}$} 
\newcommand{\MB}{$M_\mathrm{B}$$=0.88_{-0.12}^{+0.14}$} 
 
\newcommand{\RC}{$R_\mathrm{planet}$$=1.48\pm0.37$} 
\newcommand{\MC}{$M_\mathrm{planet}$$=2.38\pm0.26$}

\newcommand{\depthundiluted}{$\delta_{\mathrm{undil.}} = (R_\mathrm{planet}/R_\mathrm{A})^2$$=2.68\pm0.15$} 
 
\newcommand{\locxsky}{$\Delta x_\mathrm{sky}$$=0.42_{-0.43}^{+0.36}$} 
\newcommand{\locysky}{$\Delta y_\mathrm{sky}$$=0.66_{-0.35}^{+0.23}$}

\newcommand{\distancepc}{$1010_{-130}^{+150}$} 
\newcommand{\orbsepAU}{$> 500$} 
\newcommand{\orbperyr}{$> 11000$} 

\newcommand{\kms}{km\,s$^{-1}$}

\newcommand{\masy}{mas\,y$^{-1}$}

\newcommand{\mjup}{\mbox{M$_{J}$}}
\newcommand{\rjup}{\mbox{R$_{J}$}}
\newcommand{\Msun}{\mbox{$M_{\odot}$}}
\newcommand{\Rsun}{\mbox{$R_{\odot}$}}

\newcommand{\gccc}{g\,cm$^{-3}$}

\usepackage[T1]{fontenc}
\usepackage{ae,aecompl}

\title[Unmasking the hidden \Nplanet{}]{Unmasking the hidden \Nplanet{}: a hot Jupiter in an unresolved binary system}

\author[M.~N.~G{\"u}nther et al.]{
\parbox{\textwidth}{
Maximilian~N.~G{\"u}nther$^{c}$\thanks{E-mail: \href{mg719@cam.ac.uk}{mg719@cam.ac.uk}}, 
Didier~Queloz$^{c,g}$, 
Edward~Gillen$^{c}$,
Laetitia~Delrez$^{c}$,
Fran\c{c}ois~Bouchy$^{g}$,
James~McCormac$^{w, ce}$,
Barry~Smalley$^{k}$,
Yaseen~Almleaky$^{ks,ka}$,
David~J.~Armstrong$^{w, ce}$,
Daniel~Bayliss$^{w}$,
Artem~Burdanov$^{li}$,
Matthew~Burleigh$^{l}$,
Juan~Cabrera$^{d}$,
Sarah~L.~Casewell$^{l}$,
Benjamin~F.~Cooke$^{w, ce}$,
Szil\'ard~Csizmadia$^{d}$,
Elsa~Ducrot$^{li}$,
Philipp~Eigm\"uller$^{d, tu}$,
Anders~Erikson$^{d}$,
Boris~T.~G\"ansicke$^{w, ce}$,
Neale~P.~Gibson$^{q}$,
Micha\"el~Gillon$^{li}$,
Michael~R.~Goad$^{l}$,
Emmanu{\"e}l~Jehin$^{li}$,
James~S.~Jenkins$^{uc,ci}$,
Tom~Louden$^{w, ce}$,
Maximiliano~Moyano$^{a}$,
Catriona~Murray$^{c}$,
Don~Pollacco$^{w, ce}$,
Katja~Poppenhaeger$^{q}$,
Heike~Rauer$^{d, tu, fu}$,
Liam~Raynard$^{l}$,
Alexis~M.~S.~Smith$^{d}$,
Sandrine~Sohy$^{li}$,
Samantha~J.~Thompson$^{c}$,
St\'{e}phane~Udry$^{g}$,
Christopher~A.~Watson$^{q}$,
Richard~G.~West$^{w}$,
Peter~J.~Wheatley$^{w, ce}$
\\
}
\\
Affiliations are listed at the end of the paper.
}

\date{Last updated -; in original form -}

\pubyear{2017}

\begin{document}
\label{firstpage}
\pagerange{\pageref{firstpage}--\pageref{lastpage}}
\maketitle

\begin{abstract}
We present the discovery of \Nplanet{}, a hot Jupiter found transiting the primary star of an unresolved binary system. 
We develop a joint analysis of multi-colour photometry, centroids, radial velocity (RV) cross-correlation function (CCF) profiles and their bisector inverse slopes (BIS) to disentangle this three-body system.
Data from the Next Generation Transit Survey (NGTS), SPECULOOS and HARPS are analysed and modelled with our new {\scshape blendfitter} software.
We find that the binary consists of \NstarA{} (G6V-dwarf) and \NstarB{} (K1V-dwarf) at $<1$\arcsec{} separation.
\Nplanet{} orbits every \Nperiod{}~days.
The planet radius and mass are \RC{}~\rjup{} and \MC{}~\mjup{}, suggesting it is potentially inflated.
We emphasise that only combining all the information from multi-colour photometry, centroids and RV CCF profiles can resolve systems like \Ntarget{}.
Such systems cannot be disentangled from single-colour photometry and RV measurements alone. 
Importantly, the presence of a BIS correlation indicates a blend scenario, but is not sufficient to determine which star is orbited by the third body.
Moreover, even if no BIS correlation is detected, a blend scenario cannot be ruled out without further information.
The choice of methodology for calculating the BIS can influence the measured significance of its correlation.
The presented findings are crucial to consider for wide-field transit surveys, which require wide CCD pixels ($>5$\arcsec) and are prone to contamination by blended objects.
With TESS on the horizon, it is pivotal for the candidate vetting to incorporate all available follow-up information from multi-colour photometry and RV CCF profiles.
\end{abstract}


\clearpage
\section{Introduction}
\label{s:Introduction}

To date, more than 3700 exoplanets have been found, 2800 of which with the transit technique\footnote{\url{http://exoplanetarchive.ipac.caltech.edu/}, \mbox{online 9 March 2018}}.
Out of these, we currently know 88 (24) extra-solar binary systems (multiple systems), which contain a total of 125 (34) exoplanets\footnote{\url{http://www.univie.ac.at/adg/schwarz/multiple.html}, online 9 March 2018} \citep{Schwarz2016}.
The Next Generation Transit Survey \citep[\NGTS{};][]{Wheatley2017} and the upcoming \TESS{} mission \citep{Ricker2014} will soon further increase the sample of small planets orbiting bright stars, delivering prime targets for follow up studies.
Naturally, such wide-field \mbox{exoplanet} surveys require wide CCD pixels ($>5$\arcsec).
This can influence the observation in two ways:
1) circa $44$~per-cent of main sequence F6-K3 systems \citep{Raghavan2010} and $20-50$~per-cent of late K and M dwarfs \citep{Ward-Duong2015, Fischer1992} are actually binary and triple systems. A given target might hence be a multi-star system, whose companions remain unresolved.
2) A single CCD pixel often contains multiple background objects, whose light (and signals) influence the observations.
Both scenarios can lead to the underestimation of planet radii or to false positives \citep[see e.g.][]{Cameron2012}. 
The most common false positives are unresolved eclipsing binaries (EBs) with grazing eclipses or low-mass companions, which both can cause a shallow, planet-like transit signal. 
Another class are background eclipsing binaries (BEBs). These are faint and distant EBs aligned along the line of sight of a bright target star. This dilutes their signal onto a planetary scale.

False positives typically outnumber the planet yield by a factor of 100 \citep[see e.g.][]{Almenara2009, Latham2009, Hartman2011}.
We previously predicted for \NGTS{} that initially $\sim$5600 such false positives will outnumber the yield of $\sim$300 new exoplanets \citep{Guenther2017a}.
A series of sophisticated vetting tools have recently been developed for identifying blend scenarios and disentangling planets from false positives \citep[see e.g.][Armstrong et al., submitted]{Torres2010b,Morton2012,Diaz2014,McCauliff2015,Santerne2015a,Torres2015,Coughlin2016,Guenther2017b}.  

In this paper we evaluate an interesting signal observed with \NGTS{}, that initially seemed to originate from the transit of a hot Jupiter around a Sun-like star. After gathering HARPS follow-up spectroscopy, a planet-like radial velocity signal was confirmed, but a bisector correlation was detected. Usually, bisector correlations were seen as indicators of background eclipsing binaries, and as such the system was nearly disregarded as a false positive. Through careful analysis of all data and false positive scenarios and development of a new routine, our {\scshape blendfitter} modelling toolbox, we are able to disentangle this system.

We here present the discovery of \Nplanet{}, a hot Jupiter found orbiting a star in a still visually unresolved binary system. 
This paper attempts to provide a comprehensive case study to unmask an unresolved three-body system by combining all information from multi-colour photometry, centroids, radial velocity measurements and their bisectors.
This study is based on data gathered with the Next Generation Transit Survey (NGTS), SPECULOOS (Search for habitable Planets EClipsing ULtra-cOOl Stars, in commissioning; \citealt{burdanov2017} and Gillon et al., in prep.) and HARPS \citep{Mayor2003}, and enhanced by our recent advances with the centroiding technique for NGTS \citep{Guenther2017b}.
We here develop a new routine, {\scshape blendfitter} to conjointly model multi-colour photometry, centroids and the radial velocity (RV) extraction process. 
For this, we simulate the RV cross-correlation functions (CCFs) and study correlations of the bisector inverse span (BIS).
Our study highlights the value of a thorough inspection and modelling of multi-colour photometry, centroids, RV CCFs and BISs for exoplanet surveys.

\section{Observations}
\label{s:obs}

\Ntarget{} (\NGTSID{}; see Tab.~\ref{tab:stellar_properties}) was photometrically discovered by \NGTS{}, and followed up using high precision photometry from SPECULOOS during its commissioning period, and spectroscopy from HARPS. We detail all of these observations in this Section and provide a summary in Table~\ref{tab:obs_table}.

\begin{table}
	\centering
	\caption{Summary of all observations of \Ntarget{} used in this work, including the discovery photometry, the follow-up photometry and the spectroscopic observations.}
	\label{tab:obs_table}
	\begin{tabular}{lll}
    \hline
    Facility & Date & Notes \\
    \hline
	NGTS & 2016 Aug 18 - & 78572 points \\
     & 2017 Dec 6 & 10s exp. \\
    SPECULOOS-Callisto & 2018 Jan 26 & 301 points \\
    & & r' - 30s exp. \\
    SPECULOOS-Io & 2018 Feb 9 & 471 points \\
    & & i'+z' - 30s exp. \\
    SPECULOOS-Europa & 2018 Feb 9 & 457 points \\
    & & i'+z' - 30s exp. \\
    SPECULOOS-Callisto & 2018 Feb 15 & 445 points \\
    & & g' - 35s exp. \\
    SPECULOOS-Europa & 2018 Feb 15 & 469 points \\
    & & r' - 30s exp.\\
    HARPS & 2017 Feb 1 - & 7 spectra\\
     & 2017 Mar 5 & \\
	\hline
	\end{tabular}
\end{table}

\subsection{\NGTS{} photometry}
\label{ss:NGTS photometry}

\NGTS{} is a fully-robotised array of twelve 20\,cm Newtonian telescopes based at ESO's Paranal Observatory in Chile. The telescopes are equipped with 2K$\times$2K e2V deep-depleted Andor IKon-L CCD cameras with 13.5~$\mu$m pixels, corresponding to an on-sky size of 4.97\arcsec.

The presented data on \Ntarget{} was observed on a single \NGTS{} telescope over a photometric campaign conducted between 18 August 2016 and 6 December 2016, and detrended with the `TEST18' pipeline version. This contains 78572 exposures of 10\,s in the \NGTS{} bandpass (550 -- 927\,nm) over a total of 89 observation nights. The telescope was autoguided using an improved version of the DONUTS autoguiding algorithm \citep{McCormac2013}. The RMS of the field tracking errors was $0.136$ pixels over the 89 nights. This slightly elevated RMS (compared to the typical value of $\sim0.05$ pixels) was due to a mechanical issue with the right ascension bearing in the mount, whereby the telescope occasionally jumped by $\sim1$ pixel. The autoguiding then recentered the field after few exposures.

Image reduction, aperture photometry, and reduction of systematic effects were performed with the \NGTS{} data pipelines described in \citet{Wheatley2017}. These are based on implementations of the CASUTools\footnote{\url{http://casu.ast.cam.ac.uk/surveys-projects/software-release}, online 9 March 2018} and SysRem packages \citep{Tamuz2005}.
Light curves were screened for transit-like signals using ORION, an implementation of the box-fitting least squares (BLS) method \citep{Kovacs2002}. 
We further extracted and reduced the flux centroids of \Ntarget{} as described in \citet{Guenther2017b}. A centroid shift correlated to a transit-like signal is an indicator for contamination by a fainter background source.

\Ntarget{}'s transit-like signal of $2$~per-cent was detected with a period of \Nperiod{}~days and width of $2$~hours. No centroid shift was detected. Initially, these photometric observations alone made \Ntarget{} a strong hot Jupiter candidate.

\begin{figure*}
\centering
	\includegraphics[width=\textwidth]{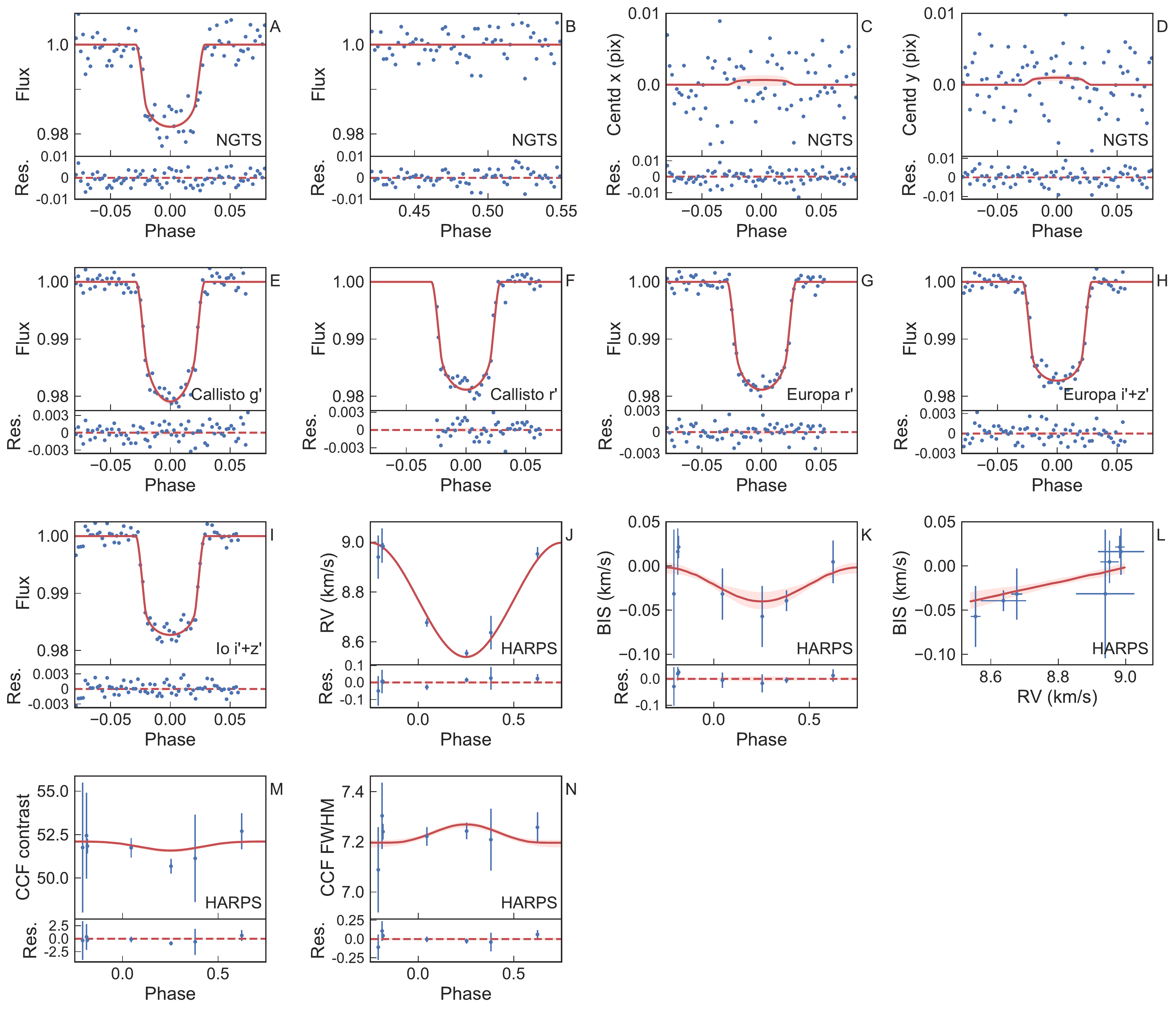}
    \caption{Data for \Ntarget{}, phase-folded at the best-fitting period of \Nperiod\,d. 
    A) \NGTS{} light curve, 
    B) \NGTS{} light curve around phase 0.5, 
    C) \NGTS{} centroid in x, 
    D) \NGTS{} centroid in y,
    E) \SPECULOOS{} Callisto g'-band,
    F) \SPECULOOS{} Callisto r'-band,
    G) \SPECULOOS{} Europa r'-band,
    H) \SPECULOOS{} Europa i'+z'-band,
    I) \SPECULOOS{} Io i'+z'-band,
    J) \HARPS{} radial velocity (RV) measurements, 
    K) \HARPS{} bisector inverse slope (BIS), 
    L) \HARPS{} BIS versus RV, 
    M) \HARPS{} Contrast measurements, 
    and N) \HARPS{} FWHM measurements.
Photometric measurements are binned equally in phase with a spacing of 0.002 (total of 500 phase-folded points).
We randomly draw 100 samples from the MCMC chain and calculate the models. Red curves in A)-N) display the median and 16th / 84th percentile of all drawn models. The global, joint modelling is described in Section~\ref{ss:Global MCMC model}.} 
    \label{fig:NOI-101129}
\end{figure*}

Table~\ref{tab:NGTS_data} provides the full photometry and centroid time series after detrending. Figure~\ref{fig:NOI-101129} shows this data phase-folded at the best-fitting transit period as determined via our global modelling (outlined in Section~\ref{ss:Global MCMC model}).

\begin{table}
	\centering
	\caption{\NGTS{} photometry and centroid data for \Ntarget{}.  The full table is available in a machine-readable format from the online journal. For guidance, ten observations are shown here.}
	\label{tab:NGTS_data}
	\begin{tabular}{cccc}
    \hline
	Time	&	Flux  & Centd x & Centd y\\
    days & (normalised) & pixel & pixel \\
    (HJD-2450000) &	& & \\
	\hline
         ...        & ...        &   ...    &   ...   \\
    7619.901516 & 1.021527545 & -0.11709990 & 0.06187227 \\
    7619.901667 & 1.000179888 & -0.04072431 & 0.04446441 \\
    7619.901806 & 0.957097368 & -0.02046733 & 0.04210692 \\
    7619.901956 & 1.076526278 & 0.07883140 & 0.03817588 \\
    7619.902106 & 0.996836033 & -0.03235835 & 0.03558102 \\
    7619.902257 & 1.123472365 & 0.10736324 & 0.00703842 \\
    7619.902419 & 1.010499832 & 0.09472378 & -0.01132131 \\
    7619.902569 & 0.943342956 & -0.06200864 & 0.05012148 \\
    7619.90272 & 1.019069713 & -0.00554865 & -0.03038287 \\
    7619.90287 & 0.961933312 & 0.03336356 & -0.09503899 \\
         ...        & ...        &   ...    &   ...   \\
	\hline
	\end{tabular}
\end{table}

\subsection{\SPECULOOS{} photometry}
\label{ss:SPECULOOS photometry}

SPECULOOS (\citealt{burdanov2017}; Gillon et al., in prep.) is located at ESO's Paranal Observatory in Chile and currently undergoing commissioning. 
The facility consists of four robotic 1-meter Ritchey-Chretien telescopes. Each telescope is equipped with an Andor Peltier-cooled deeply depleted 2K$\times$2K CCD camera with a 13.5~$\mu$m pixel size. The field of view of each telescope is 12\arcmin$\times$12\arcmin (0.35\arcsec/pixel), with optimal sensitivity in the near-infrared (700 to 1000 nm).

We observed \Ntarget{} in the g', r' and i'+z' bands during the commissioning of the first three SPECULOOS telescopes, Europa, Io and Callisto. A summary of these observations is provided in Table~\ref{tab:obs_table}. The images were calibrated using standard procedures (bias, dark, and flat-field correction) and photometry was extracted using the \textsc{IRAF/DAOPHOT} aperture photometry software \citep{stetson1987}, as described by \citet{gillon2013}. For each observation, a careful selection of both the photometric aperture size and stable comparison stars was performed manually to obtain the most accurate differential light curve of \Ntarget{}. Table~\ref{tab:SPECULOOS_data} provides the full photometry of one of the observations as an example. Figures~\ref{fig:NOI-101129}E-I show the data with the best fit determined via our global modelling (see Section~\ref{ss:Global MCMC model}).

\begin{table}
	\centering
	\caption{\SPECULOOS{} Callisto r' band photometry for \Ntarget{}.  The full table, and tables for the remaining SPECULOOS observations with Europa, Io and Callisto, are available in a machine-readable format from the online journal. For guidance, ten observations are shown here.}
	\label{tab:SPECULOOS_data}
	\begin{tabular}{ccc}
    \hline
	Time	&	Flux & Flux error \\
    days & (normalised) & (normalised) \\
    (HJD-2450000) &	& \\
	\hline
    8144.51886 & 0.99634245 & 0.00306679 \\
    8144.51931 & 0.99873645 & 0.00303029 \\
    8144.51976 & 0.9895214 & 0.00292396 \\
    8144.52021 & 0.99279671 & 0.0029041 \\
    8144.52067 & 0.99233135 & 0.00286985 \\
    8144.52112 & 0.99131786 & 0.00286618 \\
    8144.52157 & 0.98893842 & 0.00277872 \\
    8144.52202 & 0.99065349 & 0.00284851 \\
    8144.52247 & 0.98691918 & 0.00285377 \\
    8144.52292 & 0.98281773 & 0.00297856 \\
    ... & ... & ... \\
	\hline
	\end{tabular}
\end{table}

\subsection{HARPS spectroscopy}
\label{ss:Spectroscopy}

We obtained RV follow-up for \Ntarget{} with HARPS \citep{Mayor2003} on the ESO 3.6\,m telescope at La Silla Observatory in Chile between 1 February 2017 and 5 March 2017. Data were reduced using the standard HARPS reduction pipeline.
RVs were calculated for each epoch via cross-correlation of the HARPS data reduction pipeline with a G2 mask. Results along with their associated error, full width at half maximum (FWHM), contrast, and bisector slope are listed in Tab.~\ref{tab:RV}.
Early RV results were encouraging, with an in-phase variation of K$\approx$230~m\,s$^{-1}$ at a very high significance (see Fig.~\ref{fig:NOI-101129}E). However, the bisector span of the RV cross-correlation function showed a strong correlation with the measured radial velocity (see Fig.~\ref{fig:NOI-101129}F-G). 
This can often be a sign of a contaminating spectrum with large RV shifts (e.g. due to a blended binary), which is responsible for the apparent RV variation of the target \citep[][see section~\ref{ss:Bisector model}]{Santos2002}.

\begin{table}
  \scriptsize
  \centering
  \caption{HARPS radial velocities for \Ntarget{} as retrieved by the standard pipeline (DRS). The full table is available in a machine-readable format from the online journal.}
      \begin{tabular}{cccccc}
      \hline
      Time & RV	& RV error & 	FWHM	& Contrast	& BIS\\
	days	& km/s	& km/s& 	km/s& 	per-cent	& km/s\\
	HJD-2450000	& 	& 	& 	& 	& \\
    \hline
7785.721175	& 8.98228	& 0.01635	& 7.23903	& 52.138& 	0.02101\\
7790.705903	& 	8.93196& 	0.02892& 	7.00693	& 51.672& 	-0.00334\\
7791.692363	& 8.62082	& 0.01606	& 7.08774	& 50.959& 	-0.04371\\
7811.584627	& 	8.55463	& 0.01448	& 7.2421	& 51.021& 	-0.05955\\
7814.586319	& 	8.67687	& 0.01237	& 7.22015	& 52.134& 	-0.01864\\
7815.555984	& 	8.94451	& 0.01069	& 7.22257	& 52.785& 0.01179\\
7817.545532	& 	8.98783	& 0.01712	& 7.24024	& 52.212& 	0.01667\\
\hline
    \end{tabular}%
  \label{tab:RV}%
\end{table}%

\section{Analysis}
\label{s:Analysis}

\subsection{Stellar properties}
\label{ss:Stellar properties}
The \Ntarget{} system is located at \mbox{RA = 06h 17m 46.8s}, \mbox{DEC = -35d 42m 22.3s}, and is identified as \NGTSID{}, \TWOMASSID{} and \GaiaID{}, with magnitudes $G=14.4$, $J=13.3$, $K=12.8$ (Tab.~\ref{tab:stellar_properties}). 

When analysing the HARPS data we find a clear bisector correlation (Fig.~\ref{fig:NOI-101129}F-G). 
A positive correlation is a direct indicator for contamination of the spectrum of \NstarA{} by at least one other stellar object in the system (see Section~\ref{ss:Bisector model}). 
We perform a spectral fit of the seven obtained HARPS spectra to determine the parameters of the brightest object in the aperture, which we denote as \NstarA{} (Tab.~\ref{tab:stellar_properties}). 
The overall signal-to-noise ratio is relatively low (23:1), leading to large uncertainties on the derived parameters. The co-added spectrum shows no sign of contamination due to the other star in the aperture.
Using methods similar to those described by \cite{Doyle2013}, we determined values for the stellar effective temperature $T_\mathrm{eff,A}$, surface gravity $\log g_\mathrm{A}$, the stellar metallicity $[Fe/H]_\mathrm{A}$, and the projected stellar rotational velocity $(v \sin{i})_\mathrm{A}$. To constrain the latter we obtained a macroturbulence value of 2.7~km\,s$^{-1}$ using the \cite{Doyle2014} astereoseimic calibration.
We find that the effective temperature of $T_\mathrm{eff,A} = 5600 \pm 150$~K from the spectra analysis, is consistent with our results using the infrared flux method (IRFM). Lithium is not seen in the spectra, giving an upper-limit of $\log A({\rm Li})_\mathrm{A} < 1.1$.
We conclude from the measured $T_\mathrm{eff,A}$ that \NstarA{} is most likely a G6V dwarf, but consistent with a G2V to G8V dwarf \citep[see e.g.][]{Pecaut2013}.

\begin{table}
	\centering
	\caption{Stellar Properties for the \Ntarget{} system}
	\begin{tabular}{lcc}
    \hline
	Property	&	Value		&Source\\
	\hline
    \multicolumn{3}{l}{Astrometric properties of the system}\\
	\hline
    R.A.		&	94.444801			&2MASS	\\
	Dec			&	-35.706394			&2MASS	\\
    NGTS I.D.   &   \NGTSIDnr{}         &\NGTS{} \\
    2MASS I.D.	&   \TWOMASSIDnr{}	    &2MASS	\\
    \Gaia{} DR2 I.D.   &   \GaiaIDnr{}         &\Gaia{} DR2\\
    $\mu_{{\rm R.A.}}$ (\masy) & $-7.4\pm1.2$ & UCAC5 \\
	$\mu_{{\rm Dec.}}$ (\masy) & $8.6\pm1.3$ & UCAC5 \\
    \\
    \multicolumn{3}{l}{Photometric properties of the system}\\
	\hline
	V (mag)		&	$14.642\pm0.047$	&APASS\\
	B (mag)		&	$15.451\pm0.049$    &APASS\\
	g (mag)		&	$15.002\pm0.028$	&APASS\\
	r (mag)		&	$14.423\pm0.043$	&APASS\\
	i (mag)		&	$14.252\pm0.01$ 	&APASS\\
    G$_\mathrm{GAIA}$ (mag)	& $14.488$		&\Gaia{} DR2\\ 
    NGTS (mag)	& $14.109$		&This work\\
    J (mag)		&$13.281\pm0.029$		&2MASS	\\
   	H (mag)		&$12.965\pm0.029$		&2MASS	\\
	K (mag)		&$12.814\pm0.03$		&2MASS	\\
    W1 (mag)	&$12.798\pm0.023$		&WISE	\\
    W2 (mag)	&$12.820\pm0.023$		&WISE	\\
    B-V colour       & $0.809\pm0.068$       &APASS \\
    J-H colour       & $0.316\pm0.042$       &2MASS \\
    H-K colour       & $0.151\pm0.042$       &2MASS \\
    \\
    \multicolumn{3}{l}{Derived properties for \NstarA{}}\\
	\hline
    $T_\mathrm{eff,A}$ (K)    & $5600 \pm 150$               &HARPS spectra\\
    $T_\mathrm{eff,A}$ (K)    & $5570 \pm 140$               &IRFM fitting\\
    $\left[Fe/H\right]_\mathrm{A}$	 & $+0.12 \pm 0.15$		    &HARPS spectra\\
    $(v \sin{i})_\mathrm{A}$ (\kms)	     & $1.0 \pm 0.7$ 	    &HARPS spectra\\
    $\log g_\mathrm{A}$                & $4.5 \pm 0.2$			&HARPS spectra\\
    $\log A$(Li)$_\mathrm{A}$			 & $< 1.1$ 			    	&HARPS spectra\\
    $M_\mathrm{A}$ (\Msun) 		 & $1.017 \pm 0.093$  		 &ER\\
    $R_\mathrm{A}$ (\Rsun) 		 & $0.93 \pm 0.23$  		&ER\\
    $\rho_\mathrm{A}$ (\gccc) 		 & $1.09 \pm 0.29$  		& ER\\
    Spectral type, A        & G6V (G2V-G8V)           & ER2\\
	\hline
    \multicolumn{3}{l}{\begin{minipage}{3.1in}2MASS \citep{2MASS}; UCAC5 \citep{UCAC5}; APASS \citep{APASS}; WISE \citep{WISE}; \Gaia{} \citep{GAIA, Gaia2018}; ER: empirical relations using \citet{Torres2010}; ER2: empirical relations using \citet{Pecaut2013}.\end{minipage}}\\
	\end{tabular}
    \label{tab:stellar_properties}
\end{table}

\subsection{Centroiding}
\label{Centroiding}

\Ntarget{} is registered as a single source in all existing archival data. As part of the \NGTS{} candidate vetting pipeline we employ our centroiding technique \citep{Guenther2017b} to all targets. 
This test is able to detect shifts in the photometric centre-of-flux during transit events at the sub-milli-pixel level. It can identify blended eclipsing binaries at separations below 1\arcsec, well below the size of individual \NGTS\ pixels (4.97\arcsec). We previously estimated that this enables the identification of $\sim80$\% of BEBs before follow-up.

\begin{figure}
\centering
 \includegraphics[width=\columnwidth]{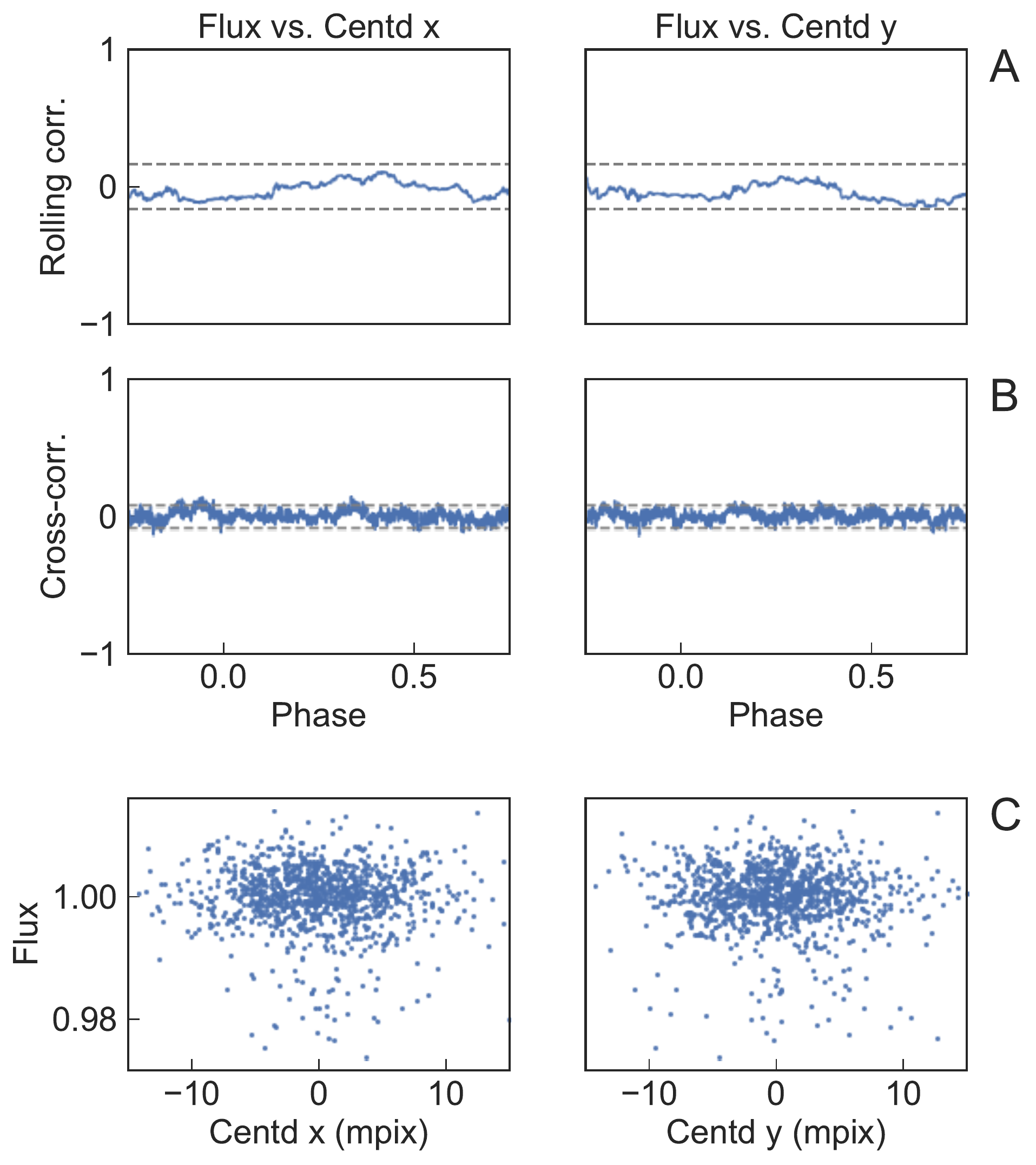}
 \caption{
 No identification of a centroid shift correlated to the transit signal for \Nplanet{}. 
The upper panels show the rolling (window) correlation (A) and cross-correlation (B) between flux and centroid, phase-folded on the best-fitting transit period.
Neither shows signs of a correlation. 
Dashed lines indicate the 99 per-cent confidence intervals in each case. 
Panel C) shows the `rain plots', a graphical illustration of the relation between flux and centroids \citep[see e.g.][]{Batalha2010, Guenther2017b}. Here, the `rain' falls straight down, meaning there is no sign of a correlation.
}
 \label{fig:Centroid_analysis}
\end{figure}

We do not observe any centroid shift for \Ntarget{} (Fig.~\ref{fig:Centroid_analysis}). 
Concurring with the \NGTS{} photometry, this initially made a planet scenario very likely.
We emphasise that the non-detection of a centroid shift minimises the risk of blends, but only completely rules out blends at more than $\sim$1\arcsec{} separation (dependent on the magnitude difference and signal depth).
In any case, the non-detection of a centroid shift allows us to place upper-limits on the possible location of this blend and the dilution it causes. 

\begin{table}
  \centering
  \caption{No statistical identification of a centroid shift in \Ntarget{}. 
The table displays the signal-to-noise ratio (SNR) of the rolling correlation and cross-correlation analyses, which are well below our threshold SNR=5 in all cases.
Further the table lists the resulting p-values from a T-test and binomial test of the in-transit centroid data, testing the Null Hypothesis that the centroid is distributed around the mean of the out-of-transit data, i.e. around $0$. All p-values are well above our threshold p=0.01 for rejecting the Null Hypothesis.}
    \begin{tabular}{l | rr}
          \hline
          & \multicolumn{1}{r}{x} & \multicolumn{1}{r}{y} \\
          \hline
SNR roll. corr.	& 1.88	& 1.35 \\
SNR cross-corr.	& 2.23	& 2.21 \\
p-value T-test	& 0.0692	& 0.1672 \\
p-value Binomial test	& 0.0649	& 0.1189 \\
    \end{tabular}%
  \label{tab:centroiding}%
\end{table}%

\subsection{HARPS CCF, RV and bisector model}
\label{ss:Bisector model}

\begin{figure}
\centering
 \includegraphics[height=0.84\textheight]{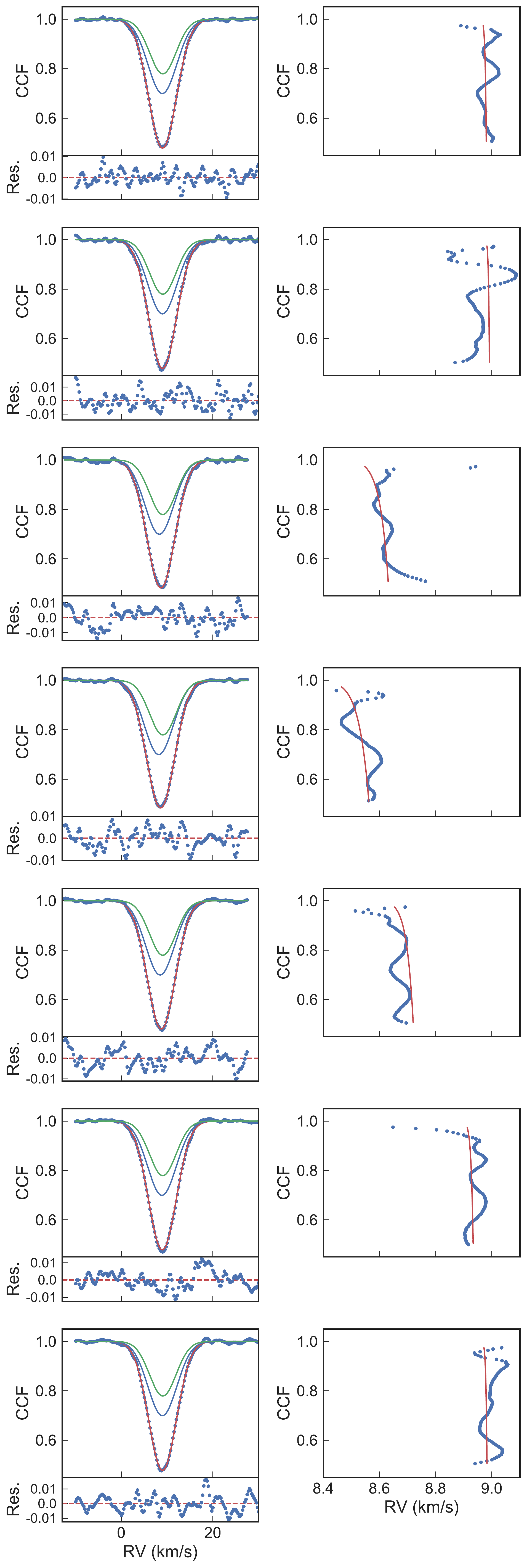}
 \caption{The seven HARPS CCF profiles (left column), and zoom onto their extracted bisectors (right column). 
 Left column: the shown CCF profiles are corrected for the best-fit baseline from the global {\scshape blendfitter} MCMC model.
 Red lines show the MCMC results for the best fit of the movement of two stars, modelled as two Gaussian profiles. The model for star A is shown in green, star B in blue and their sum in red. Sub-panels show the residuals of the fit.
 Right column: bisectors were extracted by {\scshape blendfitter} using the second derivatives of the Gaussian fit.}
 \label{fig:QL_CCFs}
\end{figure}

The radial velocity of a star is measured as the Doppler shift of spectral lines. For this, the stellar spectrum is obtained and then cross-correlated with a reference spectrum. The peak of the cross-correlation function (CCF) gives the radial velocity. In practice, it is fitted with a Gaussian function, whose mean value is the reported radial velocity (RV) value. Likewise, the full width at half maximum (FWHM) and amplitude of the Gaussian (Contrast) can be extracted.
The left column in Fig.~\ref{fig:QL_CCFs} shows the seven CCFs obtained from cross-correlating our HARPS measurements with a reference spectrum of a G2-type star (HARPS DRS has the option of a K5 and G2 mask for cross-correlations). 

The CCF bisector, in particular the bisector inverse slope (BIS), has been proven to be a powerful tool to detect star spots \citep{Queloz2001b} and background binaries \citep{Santos2002} that can mimic planet-like signals in RV data. 
The bisector is defined as the mean points halfway between equal intensities on both sides of the CCF peak. The BIS is defined as $v_t - v_b$, with $v_t$ ($v_b$) being the mean bisector velocity of all points between the top 10-40\% (the bottom 60-90\%) of the CCF peak depth \citep{Queloz2001b}.

\subsubsection{Comparison of approaches to extract the RV, FWHM and Contrast}
\label{sss:Comparison of approaches to extract the RV, FWHM and Contrast}

The most recent HARPS data reduction pipeline (HARPS DRS 3.5) fits an inverse Gaussian function with a constant baseline to the CCF profile. The RV, FWHM and Contrast measurements are then extracted as the mean, FWHM and amplitude of the Gaussian.
We implement two approaches in our {\scshape blendfitter} code.
The first choice follows the exact HARPS DRS procedure. As expected, our results match the HARPS results exactly, with a deviation of $<10^{-4}$. In all cases, this precision is by a factor of 100 within the parameters' error bars.

We find that the constant baseline approach of the HARPS DRS fit leaves strong systematic trends in the residuals of the CCF profiles. We hence implement a second method in our {\scshape blendfitter} code. Instead of using a constant baseline, we employ a Gaussian Process (GP) model jointly with our Gaussian fit and perform an MCMC fit. The MCMC and GP are implemented using {\scshape emcee} \citep{Foreman-Mackey2013} and {\scshape george} \citep{Ambikasaran2014}. 
A GP uses different kernels and metrics to evaluate the correlation between data points.
The squared distance $r^2$ between data points $x_i$ and $x_j$ is evaluated for any metric M as
\begin{align}
r^2 = ( x_i - x_j )^T M^{-1} ( x_i - x_j ).
\end{align}
In our one-dimensional case, $M$ is simplified to a scalar.
We choose our GP kernel to be
\begin{align}
k \left( r^2 \right) = c \left( 1 + 3 \sqrt{r^2} \right) e^{−3 \sqrt{r^2}},
\end{align}
which represents the product of a constant kernel $c$ and a `Matern 3/2 kernel'. This kernel can describe variations which display a rougher (i.e. more stochastic) behaviour in addition to a characteristic length scale, such as it is the case in the CCF profiles. We also fit for white noise.

We perform an MCMC fit for each CCF profile, using 50 walkers to explore the 6 dimensions (amplitude, mean, standard deviation, $c$, $M$, and a white noise scale factor). We run two separate burn-in phases of 2000 steps each, a third burn-in of 5000 steps and an evaluation of 5000 steps.
The maximum autocorrelation length for all data sets is $<100$ steps, and we hence consider all chains to be converged.
We thin the chains by a factor of 10, which leads to a total of $50 \times 5000 / 10 = 25000$ samples.

\begin{figure}
\centering
 \includegraphics[width=\columnwidth]{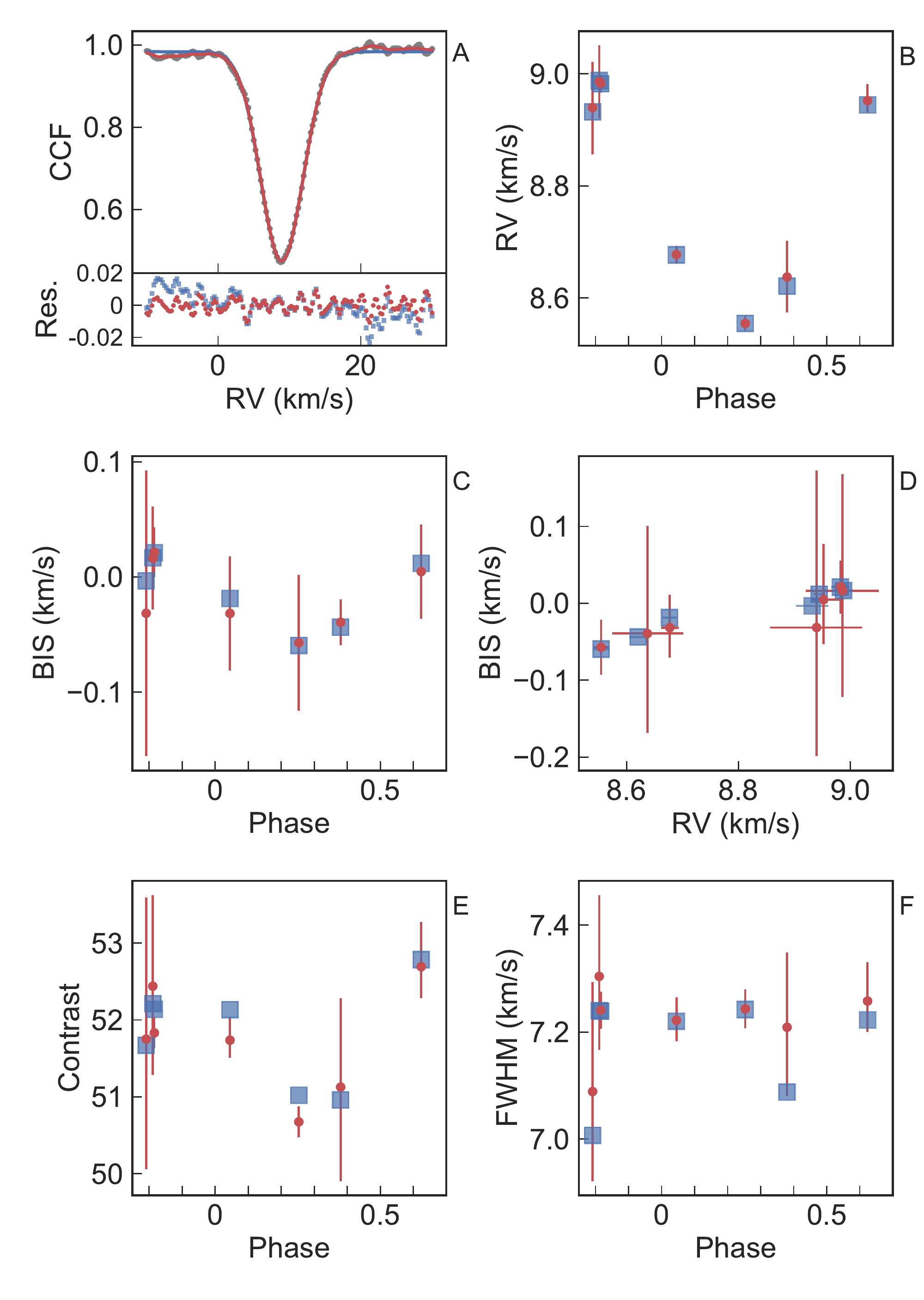}
 \caption{Comparison of the fit and residuals (A) and the extracted parameters (B-F) between the standard HARPS DRS pipeline with a constant baseline (DRS 3.5; blue squares), and our {\scshape blendfitter} code using a Gaussian Process model for the baseline (red circles). 
 The latter allows an evaluation of the parameters and error bars which is less biased due to systematic noise in the wings of the CCF profile. Values and error bars are thereby estimated with an MCMC fit and represented as the median and 16th/84th percentile of the resulting posterior likelihood distributions.}
 \label{fig:Comparison_HARPS_DRS_vs_MNG}
\end{figure}

Fig.~\ref{fig:Comparison_HARPS_DRS_vs_MNG} compares the resulting parameters from {\scshape blendfitter} and HARPS DRS. 
Reported values and error bars the median and 16th/84th percentile of the resulting posterior likelihood distributions.
The GP approach improves the fit and reduces the systematic baseline trend visible in the residuals of the HARPS DRS approach.
This shows that at the presence of strong systematics to the CCF profile, especially in the wings of the CCF profile, a constant baseline fit can be too restricting. This can lead to a high bias with low variance. The GP model allows an evaluation with lower bias and higher (`fairer') variance.
We consequently use the parameters extracted with our GP model for the global modelling in Section~\ref{ss:Global MCMC model}. The full table of these values is available in a machine-readable format from the online journal.

We here purposely use a single Gaussian model to fit the measured HARPS CCF profiles. This is to match the standard way that HARPS data is analysed (assuming a single planet model). In contrast, in our global MCMC model (see Section~\ref{ss:Global MCMC model}) we outline the detailed analysis of the HARPS CCFs with a bimodal Gaussian model (for an unresolved blended system).

\subsubsection{Comparison of approaches to extract the bisector and BIS}
\label{sss:Comparison of approaches to extract the bisector and BIS}

Throughout the literature, the CCF bisectors have been calculated in slightly different ways, three of which we outline here. 
First, the original implementation for exoplanets by \citet{Queloz2001b} builds on the approach used in studies of binary stars \citep[e.g.][]{Toner1988,Gray1989} for individual spectral lines. It uses the sampling on the left wing of the CCF peak. At each measured point a horizontal line is drawn to intersect with the right wing. The intersection value on the right wing is calculated from a linear interpolation between the two nearest points. The bisector at this level is then calculated as the mean between the left and right value.
Second, a cubic spline interpolation can be used to interpolate both sides of the CCF, and calculate the bisector at any chosen value.
Last, the most recent HARPS data reduction pipeline (HARPS DRS 3.5) further minimises the impact of outlying points. The routine fits a Gaussian function to the CCF, and calculates the line bisectors from the second derivatives of this fit.

In our {\scshape blendfitter} code, we implement these three methods of calculating the bisector: linear interpolation, cubic spline interpolation and second derivatives of a Gaussian fit. We re-analyse the HARPS CCFs to verify our implementation reproduces the reported HARPS results, and to compare the three methods with each other.
The right column in Fig.~\ref{fig:QL_CCFs} shows the extracted bisectors using the same approach as HARPS DRS. 
We note that all analysed HARPS spectra show a 'serpentine shape' in their bisectors, which can introduce systematic errors into the BIS calculation.

All three methods result in almost identical shapes of the bisectors. 
However, the linear interpolation approach leads to systematic deviations of the bisector near the top and bottom of the CCF profile.
When extracting the BIS from the bisectors, we find that for low-noise CCF profiles all three methods agreed in their BIS measurements to a few meters per second, well within their error bars. However, for high-noise CCF profiles the cubic spline solution differed from the DRS approach by up to $\sim10$~m/s, and the linear interpolation approach by up to $\sim100$~m/s. This was mainly driven by the discrepancy in extracted bisectors towards the top and bottom of the CCF profile.

We detect a BIS correlation with all three methods.
The DRS approach proves to be the most robust way to extract the BIS, while the linear interpolation is strongly affected by noise in the CCF profile. 
Our {\scshape blendfitter} software includes the choice between all three methods, but as the DRS approach proved to be the most robust, we use this setting for all following analyses.
We strongly caution that the choice of methodology for calculating the bisectors can influence the measured significance of a BIS correlation.

\subsubsection{BIS correlations: distinguishing atmospheric phenomena and blends}
\label{sss:BIS correlations: distinguishing atmospheric phenomena and blends}

If the target were a single star with no atmospheric phenomena, such as star spots, the entire CCF profile would oscillate around its mean value. Accordingly, the bisector would oscillate around its mean value, while maintaining its shape and orientation.
Two events can cause a phase-dependent trend in the BIS: changes in the stellar atmosphere \citep{Queloz2001b} and blended objects \citep{Santos2002}. 

\textit{Atmospheric phenomena}:
If a star shows strong atmospheric activity, such as star spots, the top of the RV CCF profile will remain mostly unaffected, while the bottom will show strong oscillations around the mean value. This leads to an anti-correlation between the BIS and RV measurements \citep[see e.g.][]{Queloz2001b, Boisse2011}.

\textit{Blended systems}:
If the observed target is a multiple star system whose angular separation is smaller than the fibre of the radial velocity instrument \citep[1\arcsec{} for HARPS, see][]{Mayor2003}, each obtained spectrum will show the combined blended spectra of all objects.
The measured radial velocity is the flux-weighted average of all components. In the following Section~\ref{sss:Modelling the CCFs of blended systems} we distinguish the two scenarios, whether the brighter or fainter object are orbited by a third body.

\subsubsection{Modelling the CCFs of blended systems}
\label{sss:Modelling the CCFs of blended systems}

We assume a three body system in which star A is the brightest object, star B is the second star and object C is a third body orbiting one of the stars. 
We assume the light from object C is negligible in comparison to star A and B.
We then can model the overall CCF extracted from a blended system as the sum of the CCF from star A and B. 
As the true shapes of their CCFs are unknown, we represent them as two Gaussian functions, which is a good approximation of the true shape. 
The amplitudes $A_\mathrm{A}$ and $A_\mathrm{B}$ (of the Gaussians representing star A and B) depend on the product of two factors: 1) the amount of light entering the fibre from each star, $F_\mathrm{A}$ and $F_\mathrm{B}$; 2) the intrinsic CCF contrast in dependency of the stellar spectral type, $C_\mathrm{A}$ and $C_\mathrm{B}$.
They are directly connected to the dilution for the RV data. The dilution of star B and star A are calculated as:
\begin{align}
D_\mathrm{0,B}^\mathrm{RV} &= 1 - \frac{A_\mathrm{B}}{A_\mathrm{_A} + A_\mathrm{B}} = 1 - \frac{C_\mathrm{B} F_\mathrm{B}}{C_\mathrm{B} F_\mathrm{_A} + C_\mathrm{B} F_\mathrm{B}},\\
D_\mathrm{0,A}^\mathrm{RV} &= 1 - D_\mathrm{0,B}^\mathrm{RV}
\label{eq:dil_RV}
\end{align}

We retrieve the values for $F_\mathrm{A}$ and $F_\mathrm{B}$ from our dilution model (see Section~\ref{ss:Dilution model}). 
We further study the dependency of the contrast $C_\mathrm{A}$ and $C_\mathrm{B}$ on the stellar spectral type.
\cite{Sousa2008} performed a study of 451 potential exoplanet hosts with HARPS, and estimated their effective temperatures, surface gravities and metallicities.
We retrieve the original CCFs from the HARPS archives, and extract the measured amplitudes of these targets.
The CCF contrast strongly depends on the metallicity. We assume that star B has a comparable metallicity to star A, and select only objects with Fe/H between -0.03 and 0.27 (see Tab.~\ref{tab:stellar_properties}).
We further only select objects analysed with the HARPS CCF G2 mask, to be consistent with our data set. This limits the sample to stars $\gtrapprox 5000$~K.
We note that the contrast also strongly depends on the vsini of the star. The sample from \cite{Sousa2008} only considers vsini $\lessapprox3$~km/s, and is hence biased in this regard.
Due to these sample limitations, we can not formulate an empirical relation between the CCF contrast and the stellar type for all possible parameter ranges in our global model.
Therefore, we choose to instead propagate the range of possible contrast values from 40\% to 60\% as an uncertainty onto our prior for the dilution via Eq.~\ref{eq:dil_RV}.

\begin{figure*}
    \centering
    \begin{subfigure} 
        \centering
 	    \includegraphics[width=1\linewidth]{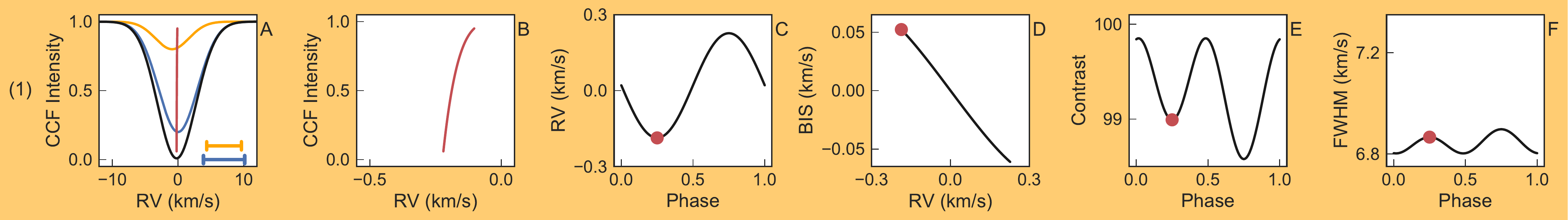}
    \end{subfigure}%
    ~ 
    \begin{subfigure}
        \centering
 		\includegraphics[width=1\linewidth]{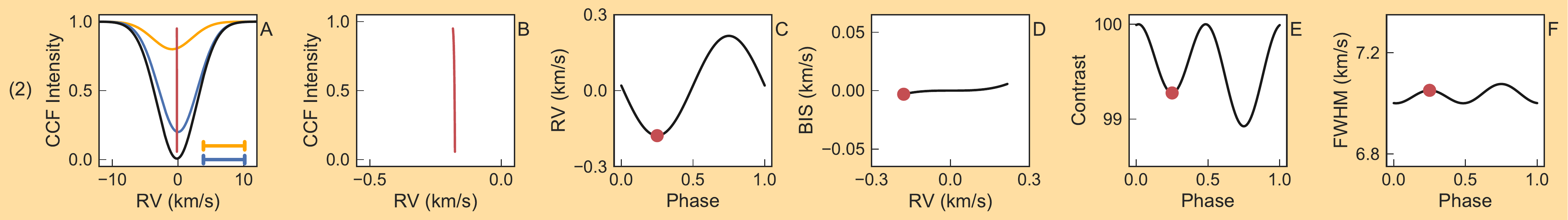}
    \end{subfigure}
    ~ 
    \begin{subfigure}
        \centering
 		\includegraphics[width=1\linewidth]{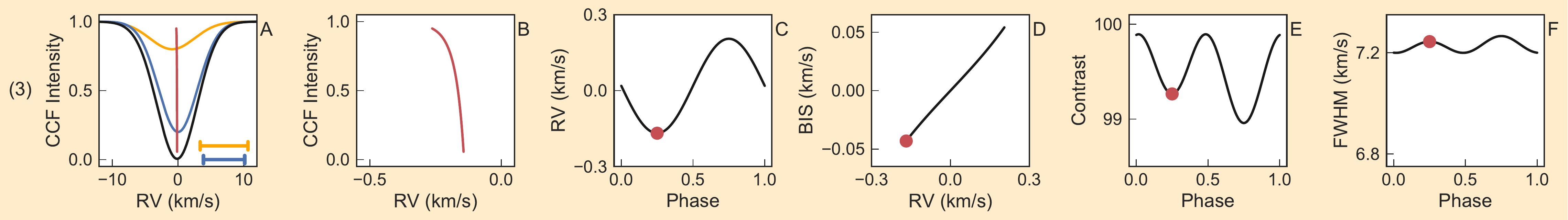}
    \end{subfigure}
    ~
        \begin{subfigure} 
        \centering
 	    \includegraphics[width=1\linewidth]{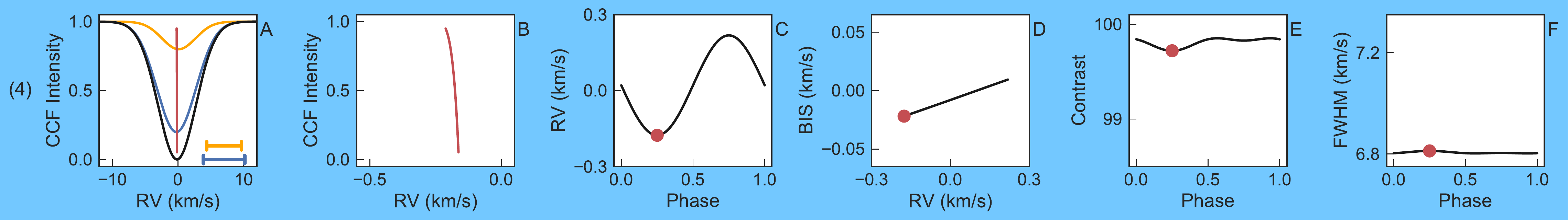}
    \end{subfigure}%
    ~ 
    \begin{subfigure}
        \centering
 		\includegraphics[width=1\linewidth]{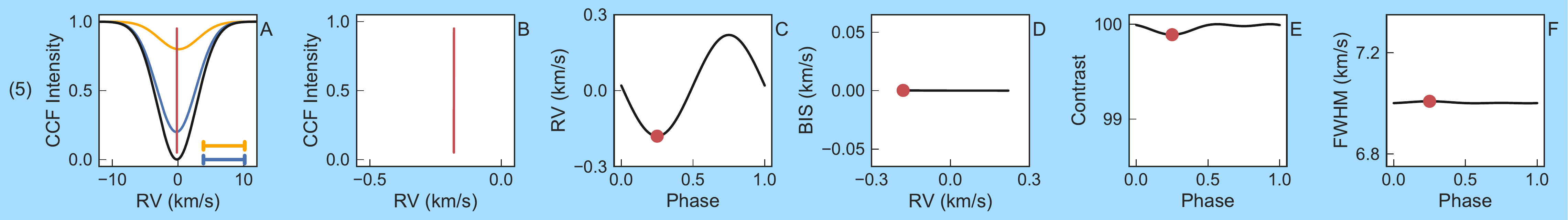}
    \end{subfigure}
    ~ 
    \begin{subfigure}
        \centering
 		\includegraphics[width=1\linewidth]{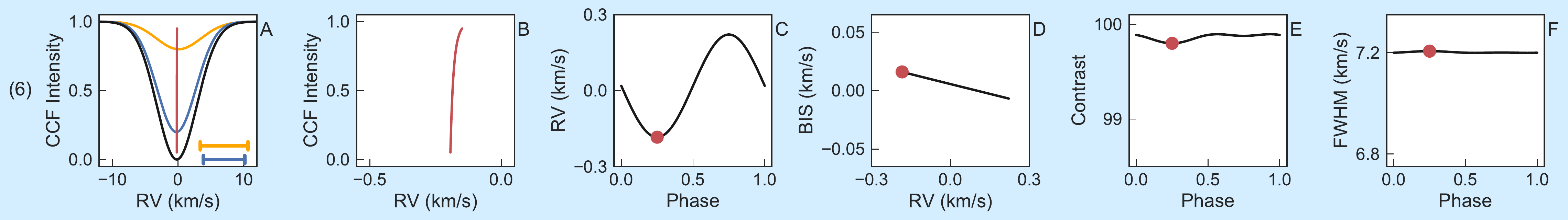}
    \end{subfigure}
 \caption{Example scenario of an unresolved binary system, where one star is orbited by a gas giant planet or brown dwarf. The primary was set to a systemic RV of $0$~km/s, the secondary to $0.1$~km/s, reflecting the orbital motion of the two binary stars.
 The numbering of the scenarios refers to Section~\ref{sss:Modelling the CCFs of blended systems}.
 The top three panels (orange background) display the scenario of a brown dwarf orbiting star B with $K=1$~km/s. The FWHM or star B varies. 
 First panel: FWHM$_\mathrm{B}$ < FWHM$_\mathrm{A}$; 
 second panel: FWHM$_\mathrm{B}$ = FWHM$_\mathrm{A}$; 
 third panel: FWHM$_\mathrm{B}$ > FWHM$_\mathrm{A}$.
 The bottom three panels (blue background) display the scenario of a gas giant planet orbiting star A with $K=0.25$~km/s. 
 The FWHM or star A varies. 
 Fourth panel: FWHM$_\mathrm{B}$ < FWHM$_\mathrm{A}$; 
 fifth panel: FWHM$_\mathrm{B}$ = FWHM$_\mathrm{A}$; 
 sixth panel: FWHM$_\mathrm{B}$ > FWHM$_\mathrm{A}$.
 A) simulated CCF profile (black) and bisector (red). The profile is modelled as the sum of two Gaussian functions representing star A (blue) and star B (orange). The horizontal lines at the bottom right indicate the ratio of the FWHM.
 B) Close-up of the bisector, measured from a single Gaussian fit. 
 C) The RV signal, measured from a single Gaussian fit, resembles a typical hot Jupiter observation in all cases. 
 D) The correlation of the BIS with the RV signal is a function of dilution, offset in systemic RV, and FWHM of the two stars.
 E) Total CCF contrast, measured from a single Gaussian fit.
 F) Total FWHM, measured from a single Gaussian fit.
 The red circles in C-F) denote at which time the snapshot shown in A) and B) was taken.
 The offset from (0,0) in D) and the different peak height in E) and F) result from the different RV zero-points of the primary and secondary.
 All measurements were extracted with our {\scshape blendfitter} tools.
 A color version and an animated version of this figure is available from the online journal.}
 \label{fig:CCF_bisector_toy_model}
\end{figure*}

Similar to the analysis by \cite{Santos2002}, we use our CCF model to investigate the effect of two blend scenarios on the RV and BIS measurements in a ``toy model". 
Fig.~\ref{fig:CCF_bisector_toy_model} displays all six simulated scenarios, which we outline in the following.

\textbf{Scenarios 1-3: star B is orbited by object C.}
We simulate two Gaussians with $D_0^\mathrm{RV}=0.8$ and RV semi-amplitude K$_\mathrm{B}=2$~km/s. FWHM$_\mathrm{A}$ is fixed at $7$~km/s, and FWHM$_\mathrm{B}$ is varied between $6.8$~km/s, $7$~km/s, and $7.2$~km/s.
We then use our {\scshape blendfitter} toolbox to extract the RV and bisector measurements.
~
\begin{enumerate}[label={(\arabic*)}]
\setcounter{enumi}{0}
\item FWHM$_\mathrm{B}$ < FWHM$_\mathrm{A}$: 
The measured BIS is anti-correlated with the RV value.
We hence caution that this scenario can mimic BIS anti-correlations introduced by atmospheric turbulence.

\item FWHM$_\mathrm{B}$ = FWHM$_\mathrm{A}$: 
In practice, the BIS correlation would be covered by noise and not be measurable.
We hence caution that blended objects with similar FWHM can remain undetected and lead to miss-classification of object C. This can lead to a wrong planet mass or false positives.

\item FWHM$_\mathrm{B}$ > FWHM$_\mathrm{A}$: 
The measured BIS is correlated with the RV value.
\end{enumerate}

\textbf{Scenarios 4-6: star A is orbited by object C.}
We simulate two Gaussians with $D_0^\mathrm{RV}=0.8$ and RV semi-amplitude K$_\mathrm{A}=0.45$~km/s. FWHM$_\mathrm{A}$ is again fixed at $7$~km/s, and FWHM$_\mathrm{B}$ varied between $6.8$~km/s, $7$~km/s, and $7.2$~km/s.
~
\begin{enumerate}[label={(\arabic*)}]
\setcounter{enumi}{3}
\item FWHM$_\mathrm{B}$ < FWHM$_\mathrm{A}$: 
The measured BIS is correlated with the RV value.

\item FWHM$_\mathrm{B}$ = FWHM$_\mathrm{A}$: 
In practice, the BIS correlation would be covered by noise and not be measurable.
We hence caution that blended objects with similar FWHM can remain undetected and lead to miss-classification of object C. This can lead to a wrong planet mass or false positives.

\item FWHM$_\mathrm{B}$ > FWHM$_\mathrm{A}$: 
The measured BIS is anti-correlated with the RV value.
We hence caution that this scenario can mimic BIS anti-correlations introduced by atmospheric turbulence.
\end{enumerate}

We emphasise that there is no difference between the extracted RV curves of all scenarios (Fig.~\ref{fig:CCF_bisector_toy_model}). 
This underlines that including a precise bisector analysis in a global model is pivotal to minimise the false positive risk for exoplanet candidates. 
If a BIS correlation is detected, the signal can still originate from either star A or star B. Disentangling such a system requires global analysis conjoint with multi-color information, as presented in the following.
However, even in cases where no bisector correlation is detected, scenarios 2 and 5 show that a blend scenario can not be ruled out without further information.

\subsubsection{Model of the CCF FWHM of \NstarA{} and \NstarB{}}
\label{sss:Model of the CCF FWHM of NstarA and NstarB}

The HARPS CCF profile's FWHM is a function of the stellar rotation and spectral type. From empirical calibrations, it can be expressed as a function of the star's $v\sin{i}$ and B-V colour: 
\begin{align}
\sigma^2 =& \left( \frac{v \sin{i}}{1.95} \right)^2 + \sigma_0^2
\end{align}
~
\begin{align}
\begin{split}
\sigma_0^2 =& \left( 8.625 - 20.037 [B-V] + 23.388 [B-V]^2 \right. \\
           & \left. - 10.364~[B-V]^3 + 1.273~[B-V]^4 \right)^2
\end{split}          
\end{align}
~
\begin{align}
\mathrm{FWHM} =& 2 \sqrt{ 2 \mathrm{ln}(2) \sigma^2 }.
\end{align}
This relation is only valid for main-sequence FGK stars with effective temperatures $T_\mathrm{eff} \gtrapprox 3900$~K

\begin{figure}
\centering
 \includegraphics[width=\columnwidth]{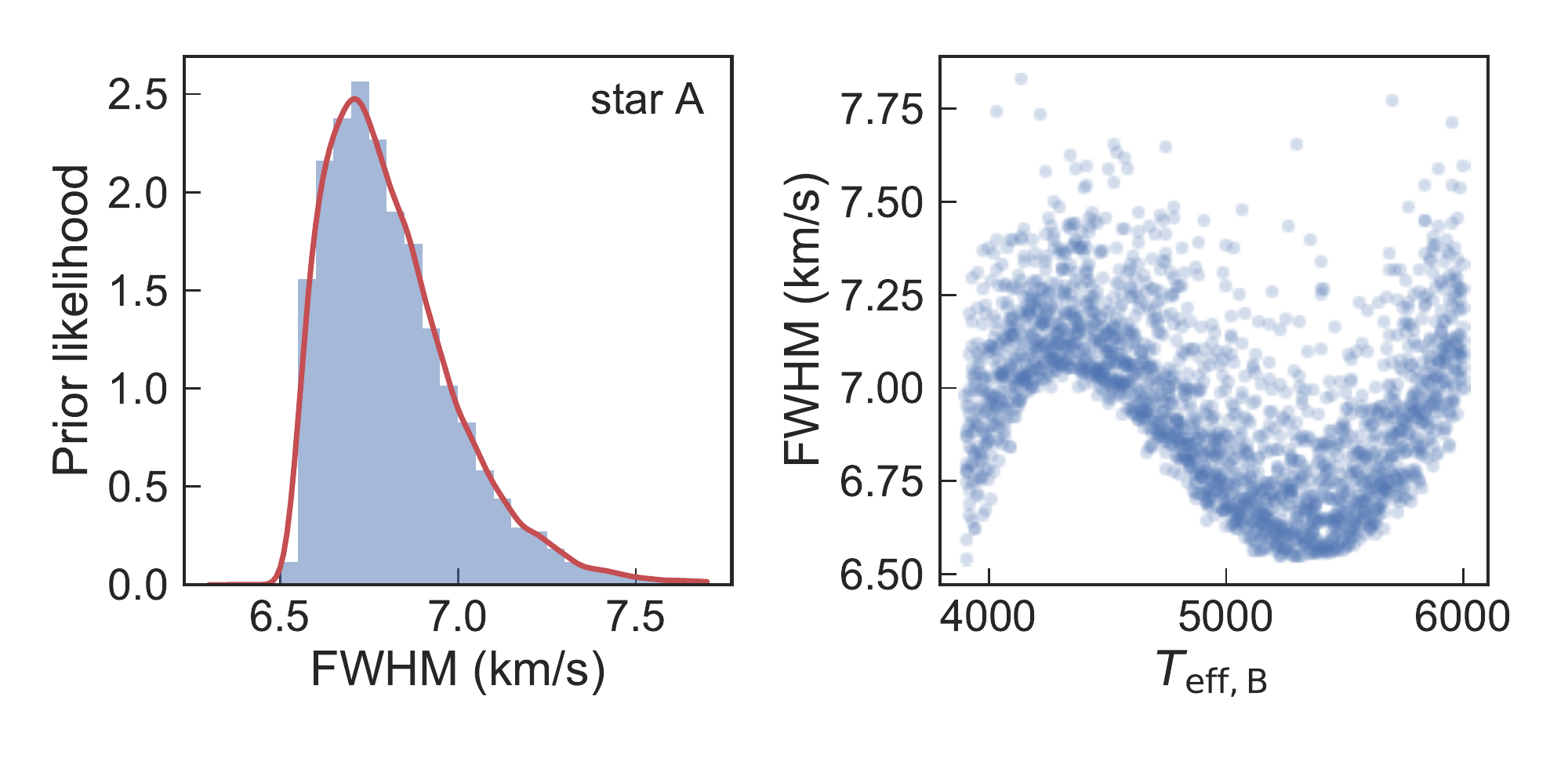}
 \caption{Prior likelihood distributions for the FWHM of star A (A) and star B (B), the latter expressed as a function of $T_\mathrm{eff,B}$. Note the minima of the FWHM relation for early K-type stars.}
 \label{fig:FWHM_Teff_relation}
\end{figure}

We next use the relations by \citet{Sekiguchi2000} to relate the B-V colour to the effective temperature $T_\mathrm{eff}$, metallicity $[Fe/H]$ and surface gravity $\log{g}$.
\begin{align}
\begin{split}
\tiny
[B-V] =& - 813.3175 + 684.4585 \log{T_\mathrm{eff}} \\
	   &- 189.923 \log{T_\mathrm{eff}}^2 + 17.40875 \log{T_\mathrm{eff}}^3 \\
       &+ 1.2136 [Fe/H] + 0.0209 [Fe/H]^2 \\
       &- 0.294 [Fe/H] \log{T_\mathrm{eff}} -1.166 \log{g} \\
       &+ 0.3125 \log{g} \log{T_\mathrm{eff}}
\end{split}
\end{align}
With the values and uncertainties for star A from the spectral analysis (see Table~\ref{tab:stellar_properties}), we use these relations to calculate a prior on the FWHM of star A (shown in Fig.~\ref{fig:FWHM_Teff_relation}A).

Next, we establish a prior on star B in dependency of $T_\mathrm{eff,B}$, which is calculated from the dilution relation (Section~\ref{ss:Dilution model}) and updated at each step in the MCMC.
We assume that both stars formed in the same system, and hence that star B has a similar metallicity to star A.
Further, as there are no signs of strong stellar line broadening, we assume that star B is a slow rotator like star A. 
We then evaluate the above relations for a range of $T_\mathrm{eff,B}$ from $3900-6000$~K in steps of $1$~K. Fig.~\ref{fig:FWHM_Teff_relation}B shows a sampling of the resulting prior on $\mathrm{FWHM_B}$.
Note the minima of the FWHM relation for early K-type stars.

\subsection{Global dilution model}
\label{ss:Dilution model}

We assume that \NstarA{} dominates the observed light, and that the spectral analysis of the HARPS data constrains the properties of \NstarA{}.
Additionally, our joint modelling of photometry and RV allows to make use of some informative priors and constraints on star B. This is incorporated in the dilution terms for star A and star B for the photometric data:
\begin{align}
D_\mathrm{0,B}^\mathrm{phot} =& 1 - \frac{F_\mathrm{B}}{F_\mathrm{A} + F_\mathrm{B}},\\
D_\mathrm{0,A}^\mathrm{phot} =& 1 - D_\mathrm{0,B}^\mathrm{phot}.
\label{eq:dilution}
\end{align}

With the knowledge of the spectral type of \NstarA{}, we can simulate the dilution originating from different stellar companions using the telescope transmission functions and stellar model spectra.
We make use of the PHOENIX stellar models \citep{Allard1995,Husser2013}. These are given in a grid, encompassing the effective temperature $T_\mathrm{eff}$ in steps of $100$~K, $\log{g}$ in steps of $0.5$, and $[Fe/H]$ in steps of 0.5 for our range of possible properties. In practice, we employ the {\scshape pysynphot} software package \citep{Lim2013}, which allows to retrieve an interpolated spectrum for any requested property.

\begin{figure}
\centering
 \includegraphics[width=\columnwidth]{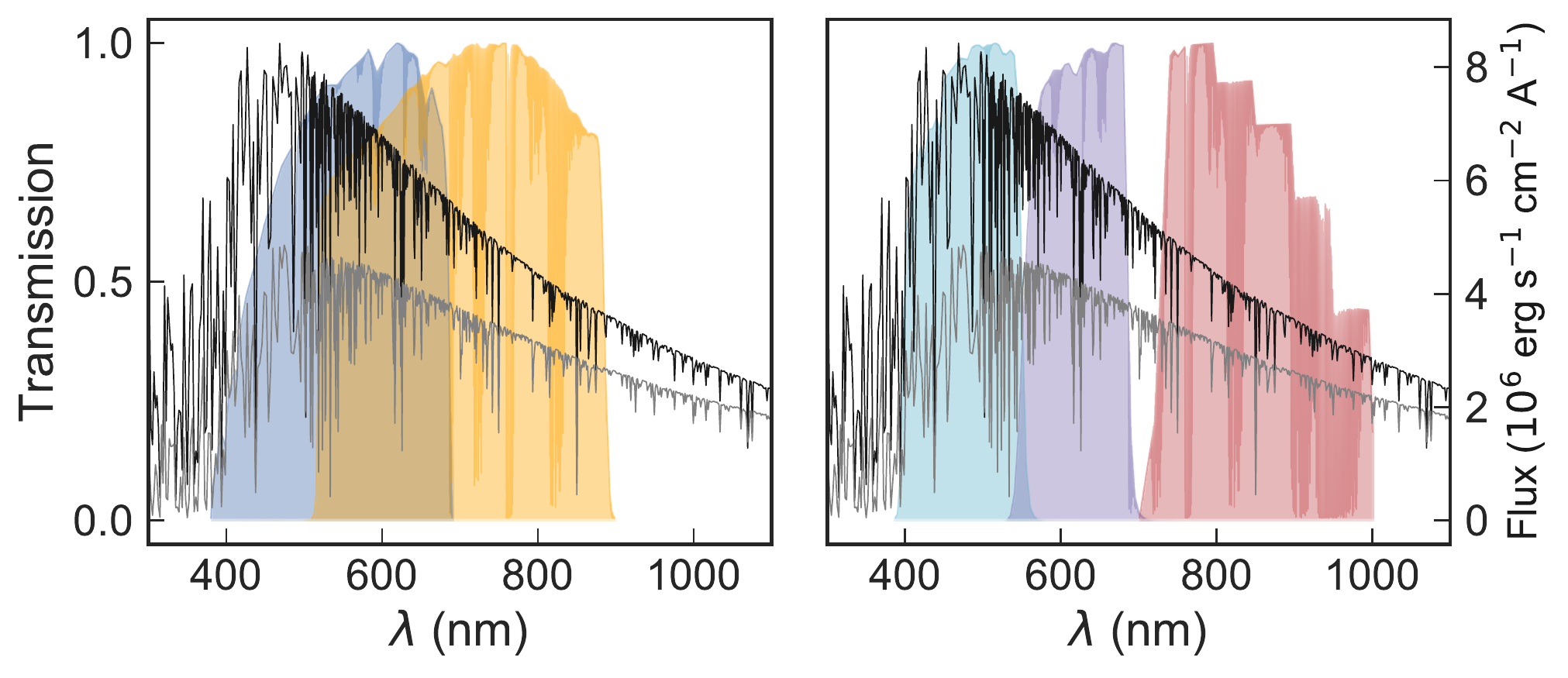}
 \caption{Dilution is a function of the instrument transmission and stellar spectral types. Left axis: Transmission efficiency of HARPS (blue), NGTS (orange) and the SPECULOOS g'-band (light blue), r'-band (purple), and i'+z'-band (red), all including atmospheric absorption. Right axis: luminosity of a G6V (top) and a K4V (bottom) star. The different bandpasses lead to a different dilution of the planetary signal for each instrument. 
 }
 \label{fig:transmission_example}
\end{figure}

We employ the transmission functions of the NGTS, SPECULOOS and HARPS instruments \citep[][private correspondence with the SPECULOOS consortium]{Wheatley2017,HARPS-user-manual}, which we multiply with a model of Earth's atmospheric absorption. Fig.~\ref{fig:transmission_example} shows all resulting transmission functions, and the model spectra of a G6V and K4V dwarf overlayed as examples. 

\begin{figure}
\centering
 \includegraphics[width=\columnwidth]{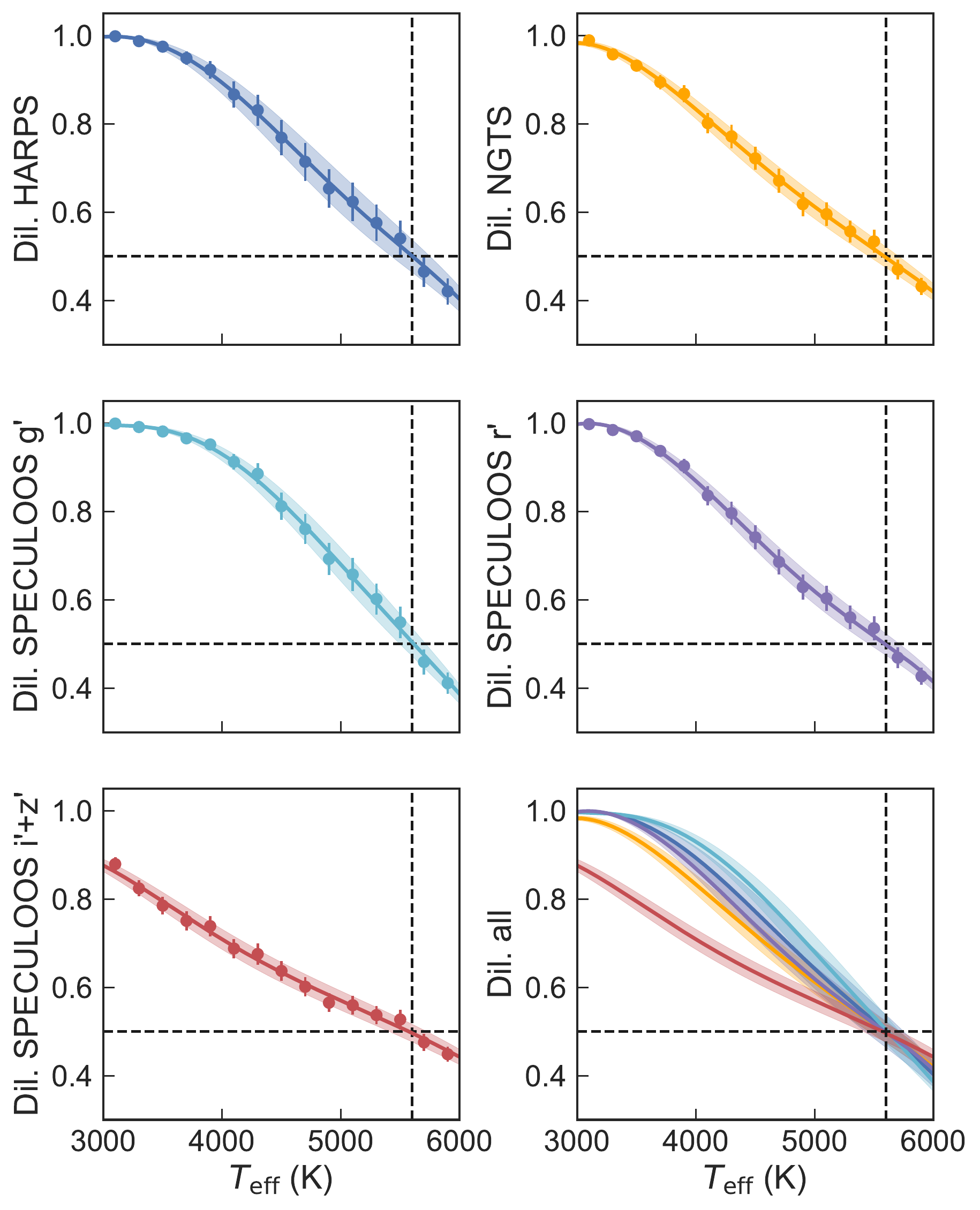}
 \caption{HARPS, NGTS and SPECULOOS dilution of star B, $D_\mathrm{0,B}$, as function of its effective temperature of \NstarB{}, $T_\mathrm{eff, B}$. We derive the dilution by passing PHOENIX model spectra through the telescope bandpasses. We fit the resulting trend with a 5th-order polynomial, which can then be used to predict the dilution for each instrument at any chosen $T_\mathrm{eff, B}$. Dashed lines at dilution 0.5 and $T_\mathrm{eff} \approx 5600$~K indicate the properties of star A. Note that, when modelling a planet on star A, the dilution of the planet signal on star A is calculated as $D_\mathrm{0,A} = 1 - D_\mathrm{0,B}$.}
 \label{fig:Dilution_simulation}
\end{figure}

We study the dilution as a function of the spectral type of \NstarB{}. We simulate \NstarA{} with the PHOENIX model for the properties (and errors) listed in Table~\ref{tab:stellar_properties}. 
Next, we simulate all possibilities for \NstarB{} by passing each PHOENIX model spectra in $T_\mathrm{eff}$ steps of $200$~K through the HARPS and NGTS transmission functions. 
From this, we calculate the dilution of star B, $D_\mathrm{0,B}$, via Eq.~\ref{eq:dilution} as a function of the effective temperature of \NstarB{}, $T_\mathrm{eff,B}$. 
When modelling a planet on star A, the dilution of the planet signal on star A is calculated as $D_\mathrm{0,A} = 1 - D_\mathrm{0,B}$.
Fig.~\ref{fig:Dilution_simulation} shows the resulting dilution as function of $T_\mathrm{eff,B}$ for HARPS, NGTS and all used SPECULOOS filters. We perform a 5th-order polynomial fit to all mean points and errorbars. This fit can then be used to predict the dilution and its error at any chosen $T_\mathrm{eff,B}$.

\subsection{Inferring properties of \NstarB{}}
\label{ss:Inferring NstarB}

Without visual information on \NstarB{}, we have no a-priori knowledge of its spectral type and properties.
It was not possible to constrain the spectrum of \NstarB{} from the \HARPS{} spectra analysis (Section~\ref{ss:Stellar properties}) nor from an SED fit without prior information.
However, we can employ our global MCMC model of the photometric and RV data to estimate the effective temperature of \NstarB{}, $T_\mathrm{eff,B}$, from our dilution model (Section~\ref{ss:Dilution model}).
At each step in the MCMC chain of the global modelling, we sample the dilution values for all instruments.
We pass them into the dilution model, allowing us to sample the likelihood distribution of $T_\mathrm{eff,B}$. We can then employ empirical relations to use $T_\mathrm{eff,B}$ for estimating the likelihood distribution of the radius $R_\mathrm{B}$ and mass $R_\mathrm{B}$ (see Sections~\ref{ss:Model of the RV between NstarA and NstarB} and \ref{ss:Identifying NstarB}).
In inferring properties of \NstarB{} we make the assumption that it is a main sequence star. A giant star would dominate the light and would have been identified in the \HARPS{} spectra analysis. Moreover, low mass main sequence stars are the most abundant objects in the night sky, and frequent companions in binary systems with a G-type primary.

\subsection{Model of the RV offset between \NstarA{} and \NstarB{}}
\label{ss:Model of the RV between NstarA and NstarB}

In order to model the two CCFs, the systemic radial velocities of \NstarA{} and \NstarB{} are needed. We will use both as free parameters to the fit in Section~\ref{ss:Global MCMC model}. However, they are tied to each other by astrophysical constraints, which we can calculate and include into our MCMC modelling.

We can calculate the RV semi-amplitude for each star, $K_\mathrm{A,B}$, in the binary system as
\begin{equation}
K_\mathrm{A,B} = \frac{M_\mathrm{B,A} \sin{i_\mathrm{binary}}}{(M_\mathrm{A} + M_\mathrm{B})^{2/3}} 
~\frac{ (2 \pi G)^{1/3} }{P_\mathrm{binary}^{1/3} (1-e_\mathrm{binary}^2)^{1/2}}.
\label{eq:K}
\end{equation}

Here, $M_\mathrm{A}$ and $M_\mathrm{B}$ are the masses of \NstarA{} and \NstarB{}, respectively. $P_\mathrm{binary}$, $i_\mathrm{binary}$ and $e_\mathrm{binary}$ are the period, inclination and eccentricity of the binary system (not to be confused with the parameters of the planet's orbits). $G$ is the gravitational constant.

As we have no prior knowledge about this binary system, we employ a series of empirical relations to sample the likelihood space for $K_\mathrm{A,B}$ using a Monte Carlo approach.
We use our result for $M_\mathrm{A}$ as a normal prior on this parameter (see Tab.~\ref{tab:stellar_properties}).
The inclination $i\mathrm{binary}$ is randomly drawn from a uniform distribution in $\cos{i_\mathrm{binary}}$ between 0 and 90 degree.
The logarithm of the period $P_\mathrm{binary}$ is randomly drawn from a normal distribution with mean 5.03 and standard deviation 2.28 \citep{Raghavan2010}. 
The eccentricity $e_\mathrm{binary}$ is randomly drawn from the results of \citep{Tokovinin2016}. We do not use their linear fit solution, but instead calculate an empirical cumulative distribution function (CDF) of $e_\mathrm{binary}$ from their tabulated data. We interpolate the CDF with a cubic spline function, and perform random sampling from the inverse CDF.
In total, we generate 1000 random binary systems.

We then calculate the measured RV difference in dependency of the relative orbital position of the binary system, using
\begin{align}
\mathrm{RV_{A,B}}(t) &= K_\mathrm{A,B} (\cos(\nu(t) + \omega_\mathrm{A,B}) + e \cos(\nu(t))),\\
\mathrm{\Delta RV}(t) &= \abs{\mathrm{RV_{A}}(t)-\mathrm{RV_{B}}(t)}.
\label{eq:deltaRV}
\end{align}
Here, $\mathrm{\Delta RV}$ denotes the difference in systemic RV that we expect between the two stars, which is the direct result of their gravitational pull on each other.
$\nu$ is the true anomaly of the system, and $\omega_\mathrm{A,B}$ the argument of periastron with $\omega_\mathrm{B} = \omega_\mathrm{A}-180\deg$.

The parameter $\omega_\mathrm{A}$ is sampled from a uniform distribution between 0 and 360 degree. For each system, we compute $\nu$ as a function of time. This is done by calculating the mean anomaly, and then solving Kepler's equation for the eccentric anomaly. Finally, $\nu$ is computed from the eccentric anomaly. 
We evaluate $\nu$ for 100 uniformly spaced times in the range from 0 to $P_\mathrm{binary}$, sampling the entire orbit for each system.

\begin{figure}
    \centering
 	\includegraphics[width=\columnwidth]{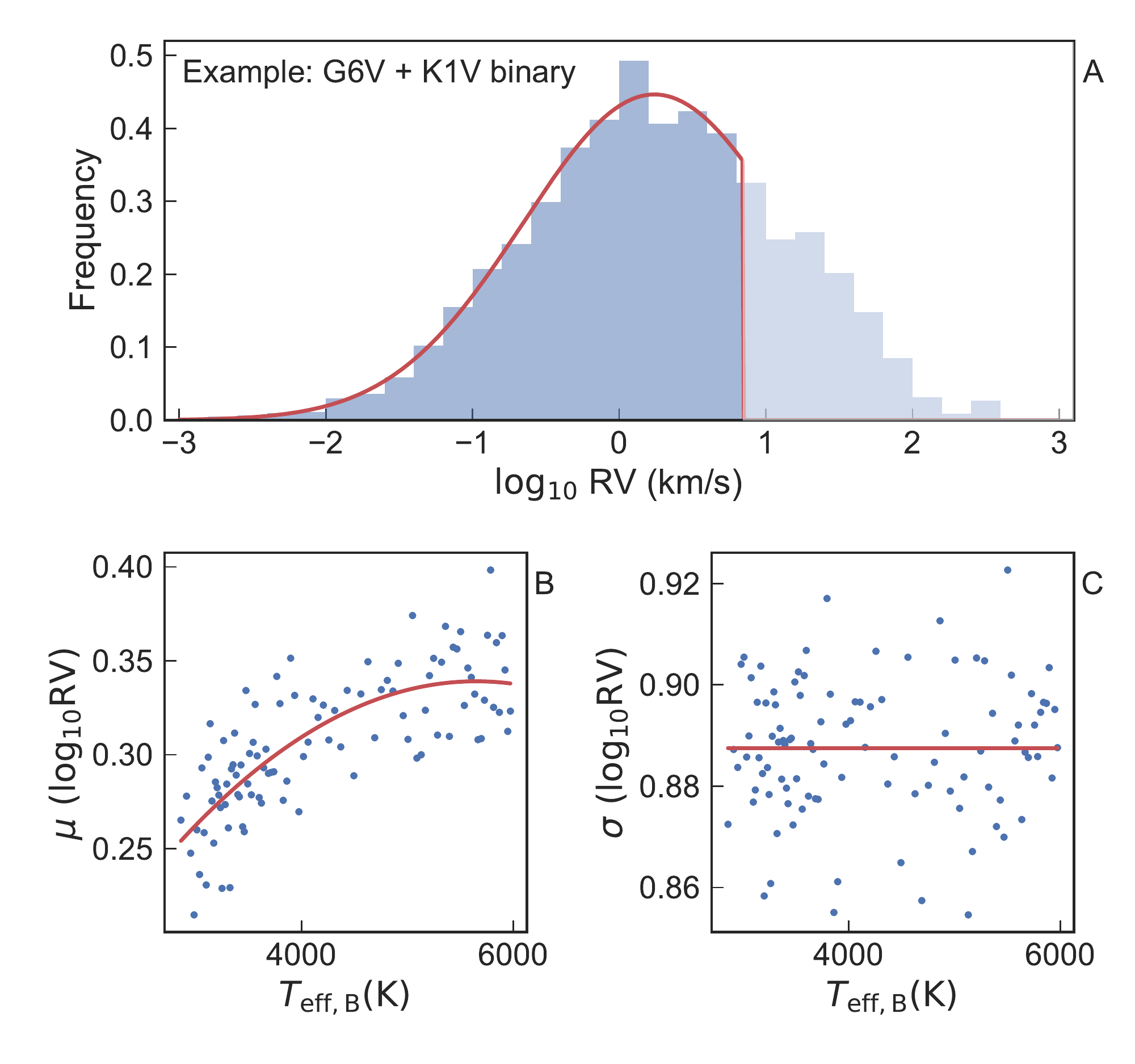}
	\caption{A) RV difference between the two stars of a G6V-K1V binary system from 1000 simulations and sampled at 100 points in phase. The truncation is set by the fact that both systems remain unresolved in HARPS. Hence, their separation has to be $\lessapprox 7$~km/s, given by the measured FWHM. The red curve shows a truncated Gaussian fit to the logarithm of the measured RV differences.
    We fix star A to the properties of \NstarA{} and simulate 1000 binary systems for $M_\mathrm{B}$ ranging from $0.1 - 1$\Msun in steps of $0.01$\Msun. We sample each system  at 100 points in phase.
   We calculate $T_\mathrm{eff,B}$ from $M_\mathrm{B}$, using our empirical relation described in Section~\ref{ss:Inferring NstarB}.
   We then calculate the mean (B) and standard deviation (C) of the Gaussian fit to $\log{\Delta RV}$ for all sampled $T_\mathrm{eff,B}$. Red curves in B) and C) show a second-order polynomial (constant) fit to the mean (standard deviation) as a function of $T_\mathrm{eff,B}$.}
 \label{fig:RV_difference}
\end{figure}

By combining all this in Eq.~\ref{eq:deltaRV}, we derive $\mathrm{\Delta RV}$ as a function of the unknown mass of \NstarB{}.
Fig.~\ref{fig:RV_difference}A shows the distribution of $\mathrm{\Delta RV}$ on the example for all simulated distributions of binary systems with a G6V primary and K1V secondary.
To generate priors for our global MCMC fit, we evaluate Eq.~\ref{eq:deltaRV} for 100 different probe masses for star B, uniformly spaced in the range 0.1-1~\Msun.
This means we have a total of $1000$ binaries $\times 100$ time points $\times 100$ probe masses $= 10^7$ samples. 

We next link mass to effective temperature using the empirical catalogue of mean dwarf stars by \citet{Pecaut2013}.
This mean dwarf model is chosen to rely on as little prior assumptions as possible for the global MCMC fit, as we initially had no information on the spectral type of \NstarB{}. It relies only on the assumption that \NstarB{} is a main sequence star (see Section~\ref{ss:Inferring NstarB}).
We use GP regression with an squared exponential kernel\footnote{also referred to as `exponentiated quadratic kernel'} and a constant kernel to fit the data in \citet{Pecaut2013}:
\begin{align}
k \left( r^2 \right) = c e^{ − \sqrt{r^2} }
\end{align}
(for discussion of the GP fitting procedure see Section~\ref{sss:Comparison of approaches to extract the RV, FWHM and Contrast}).
The resulting fit is then used to predict $T_\mathrm{eff,B}$ for any requested $M_\mathrm{B}$, translating the prior on $\mathrm{\Delta RV}$ to be a function of $T_\mathrm{eff,B}$. 
Its value is calculated at each step in the MCMC chain as described in Section~\ref{ss:Inferring NstarB}.

Next, we fit the resulting logarithmic distribution of $\mathrm{\Delta RV}$ with a Gaussian function.
When studying the mean $\mu \left( \log_{10} \mathrm{\Delta RV} \right)$ and standard deviation $\sigma \left( \log_{10} \mathrm{\Delta RV} \right)$ of this Gaussian function in dependency of $T_\mathrm{eff,B}$, we find a clear trend (Fig.~\ref{fig:RV_difference}B and C). 
We describe $\mu \left( \log_{10} \mathrm{\Delta RV} \right)$ with a second order polynomial and $\sigma \left( \log_{10} \mathrm{\Delta RV} \right)$ by its mean value. We substitute $x = (T_\mathrm{eff, B} - 3000~\mathrm{K})/3000~\mathrm{K}$, and find the following relations:
\begin{align}
\mu \left( {\log{\Delta \mathrm{RV}}} \right) &= -0.144 x^2 + 0.212 x + 0.262 \\
\sigma \left( {\log{\Delta \mathrm{RV}}} \right) &=  0.887.
\label{Eq:dRV_params}
\end{align}

These equations are then used in our MCMC model (Section~\ref{ss:Global MCMC model}) to constrain the systemic velocities in relation to each other for any evaluated $T_\mathrm{eff,B}$.
Additionally, an upper limit on $\Delta RV$ is set by the fact that both systems remain unresolved in HARPS. Hence, their separation has to be \mbox{$\lessapprox 7$~km/s}, constrained by the measured FWHM.
We hence implement a truncated Gaussian prior on $\Delta RV$ as a function of $T_\mathrm{eff,B}$.

\subsection{Detrending NGTS' photometric and centroid data with Gaussian Process regression}
\label{ss:Detrending NGTS' photometric and centroid data with Gaussian Process regression}

To decrease the influence of systematic noise, we pre-whiten the photometric and centroid data from \NGTS{}. We first mask all data during primary and secondary eclipse. We then employ a GP regression fit using the product of a Matern 3/2 kernel and a constant kernel (see also Section~\ref{sss:Comparison of approaches to extract the RV, FWHM and Contrast}). We detrend the lightcurve and centroid curves with the resulting GP.

\subsection{Global MCMC model}
\label{ss:Global MCMC model}

We perform a global, joint MCMC modelling of all data sets: the GP detrended photometric and centroid data from \NGTS{}, the HARPS cross-correlation functions (CCFs) for the seven exposures, and the extracted HARPS RV and bisector measurements.

\textbf{Priors.} 
In multiple initial MCMC test-runs we explore the scenario of a planet or sub-stellar object orbiting either star A or star B. We also explore the parameter space from different starting points, with different priors and more free parameters. We find that all approaches converge to the scenario of a planet orbiting star A.
From our previous analyses (Sections~\ref{ss:Stellar properties}-\ref{ss:Model of the RV between NstarA and NstarB}) we can hence put various priors and constraints (Tab.~\ref{tab:priors}; see Tab.~\ref{tab:MCMC_results} for a description of the parameters):
\begin{enumerate}
\item an upper limit of 1\arcsec~projected separation between \NstarA{} and \NstarB{} (acceptance of the HARPS fibre) as a uniform, informative prior constraining the centroid model. One \NGTS{} pixel spans $4.97$\arcsec, leading to limits of \mbox{$\Delta x \in (-0.2,0.2)$~pixel} and \mbox{$\Delta y \in (-0.2,0.2)$~pixel}.
\item the dilution relation from Section~\ref{ss:Dilution model}, linking the different instruments. As we model the scenario of a planet on star A (constrained by the colour difference in transit depth), we restrict the dilution further to \mbox{$D_\mathrm{0,A} \in (0,0.5)$}.
\item the RV offset relation from Section~\ref{ss:Model of the RV between NstarA and NstarB}.
\item uniform priors on all other parameters, where applicable within physical bounds, otherwise with non-restrictive bounds. 
\end{enumerate}
We note that all our priors are jointly proper, ensuring posterior propriety. None of our priors are unbounded, and the likelihood functions for all models converge to 0 as the model deviates from the data.

\begin{table}
  \centering
  \caption{Priors for the global MCMC model. Parameters are described in Tab.~\ref{tab:MCMC_results}.}
    \begin{tabular}{ll}
    \hline
    $\Delta x$ & $\mathcal{U}(-0.2,0.2)$~pixel \\
    $\Delta y$ & $\mathcal{U}(-0.2,0.2)$~pixel \\
    $D_\mathrm{0,A}$ for each instrument & see Section~\ref{ss:Dilution model}; in $(0,0.5)$ \\
    $P$ & $\mathcal{U}(0,10^{12})$~min \\
    $T_0 - 2450000~d$ & $\mathcal{U}(0,10^{12})$~min \\
    $R_\mathrm{p}/R_\mathrm{A}$ & $\mathcal{U}(0,1)$ \\
    $(R_\mathrm{A}+R_\mathrm{p})/a$ & $\mathcal{U}(0,1)$ \\
    $\cos{i}$ & $\mathcal{U}(0,1)$ \\
    $RV_\mathrm{sys,A}$ & see Section~\ref{ss:Model of the RV between NstarA and NstarB}; in $(0,100)$~km/s\\
    $RV_\mathrm{sys,B}$ & see Section~\ref{ss:Model of the RV between NstarA and NstarB}; in $(0,100)$~km/s\\
    $K$ & $\mathcal{U}(-100,100)$~km/s \\ 
    $A_\mathrm{CCF}$ & $\mathcal{U}(0,1)$ \\
    FWHM$_{\mathrm{CCF, A}}$ & $\mathcal{U}(0,100)$~km/s \\
    FWHM$_{\mathrm{CCF, B}}$ & $\mathcal{U}(0,100)$~km/s \\
    all photometric errors & $\mathcal{U}(0,1000)$~mmag \\
    all centroid errors & $\mathcal{U}(0,1000)$~mpix \\
    RV and BIS errors & $\mathcal{U}(0,1)$~km/s \\
    FWHM error & $\mathcal{U}(0,10)$~km/s \\
    Contrast error & $\mathcal{U}(0,1)$ \\
    CCF errors & $\mathcal{U}(0,10)$ \\
    \hline
    \end{tabular}
  \label{tab:priors}
\end{table}

\textbf{Fixed values.} We fix the eccentricity to $e=0$, as there is no evidence for eccentricity from the \HARPS{} RV data \citep[see e.g. discussion in][]{Anderson2012}.
The surface brightness ratio, gravitational darkening and reflectivity are also fixed to $0$, following a planet scenario.
For each bandpass, we compute quadratic limb darkening parameters for star A from the values in Tab.~\ref{tab:stellar_properties} using the open-source code by \citet{Espinoza2015} and the PHOENIX model spectra \citep{Husser2013}. To reduce free parameters in our model, we fix the limb darkening parameters $\alpha$ and $\beta$ to the values shown in Tab.~\ref{tab:limb_darkening}.

\begin{table}
  \centering 
  \caption{Limb darkening parameters for the global MCMC model.}
    \begin{tabular}{lll}
    \hline
     & $\alpha$ & $\beta$ \\
    \hline
    NGTS: & 0.4294 & $0.2019$ \\
    SPEC. g': & 0.6993 & $0.0946$ \\
    SPEC. r': & 0.4869 & $0.1927$ \\
    SPEC. i'+z': & 0.3339 & $0.2199$ \\
	\hline
    \end{tabular}
  \label{tab:limb_darkening}
\end{table}

\textbf{Baselines.} From our re-analysis of the HARPS CCFs in section~\ref{sss:Comparison of approaches to extract the RV, FWHM and Contrast}, we find that for all studied CCFs our GP model favours a simple and continuous baseline trend, which can be closely reproduced by a low-order polynomial baseline. To minimise the complexity and number of dimensions of our MCMC model, we therefore opt to use polynomial baselines instead of GPs in the global modelling. In particular, we allow a fourth order polynomial for the baseline of the HARPS CCFs, and a second order polynomial for the baseline of the SPECULOOS data. As the NGTS data covers mostly out-of-transit data, we remove any global variation using a GP regression fit beforehand (see section~\ref{ss:Detrending NGTS' photometric and centroid data with Gaussian Process regression}), and include only a constant baseline for any NGTS data in our global model. 
In fitting the baseline polynomials, we do not implement the polynomial values as jump parameters in our MCMC, but instead perform an algebraic least squares fit to the residuals of each MCMC fit at each step in the MCMC chain. This approach was proven robust and effective in multiple previous studies \citep[see e.g.][]{Gillon2012}.

\textbf{MCMC.} The MCMC is implemented using {\scshape emcee} \citep{Foreman-Mackey2013} and the {\scshape EB} binary star model \citep{Irwin2011}. We run our MCMC analysis on $37$ dimensions with $500$ walkers for $200,000$ total steps. 
$19$ of these dimensions are scaling factors for the errors of each data set.
Across all chains, we find a median (maximal) autocorrelation length of $2,400$ ($\sim3,400$) steps. The total chain is $\sim83$ ($\sim59$) times its median (maximal) autocorrelation length, which is considered as sufficient for convergence. We discard the first $\sim50,000$ steps as burn-in phase, and thin the chain by a factor of $2,500$. This results in $(200,000-50,000)/2,500*500 = 30,000$ independent samples.

\textbf{Results.} 
The hot Jupiter \Nplanet{} is orbiting \NstarA{} with a period of \Nperiod{}.
The planet radius and mass are \mbox{\RC{}~\rjup{}} and \mbox{\MC{}~\mjup{}}, conform with a potentially inflated gas giant planet.
We find a dilution of $0.38-0.43$ of the transit signal, depending on the instrument bandpass. 
\Nplanet{} has an undiluted transit depth of \mbox{\depthundiluted{}~per-cent}.
The planet introduces an undiluted RV signal of \mbox{\RVK{}~km/s} on \NstarA{}.
The systemic velocities of \NstarA{} and B are \mbox{\RVsysA{}~km/s} and \mbox{\RVsysB{}~km/s}, respectively. 
All results of our MCMC analysis can be found in Fig.~\ref{fig:NOI-101129} and \ref{fig:corner}, and Tab.~\ref{tab:MCMC_results}.

\subsection{Identifying \NstarB{}}
\label{ss:Identifying NstarB}

Using the approach outlined in Section~\ref{ss:Inferring NstarB}, we estimate the effective temperature of \NstarB{} from the dilution model, and find \mbox{\TeffB{}~K}. This places \NstarB{} most likely as an K1V dwarf (ranging G9V-K2V; see e.g. \citealt{Pecaut2013}).
From this, we calculate the final radius and mass of \NstarB{}, but deviate here from Section~\ref{ss:Model of the RV between NstarA and NstarB}.
The approach in Section~\ref{ss:Model of the RV between NstarA and NstarB} was chosen to find the mass for mean dwarf stars in dependency of $T_\mathrm{eff,B}$ as we had no prior information on \NstarB{}.
This does not allow to estimate uncertainties, particularly it is not possible to propagate uncertainties on $\log{g}_\mathrm{B}$ and $[Fe/H]_\mathrm{B}$.

For the calculation of uncertainties, we here estimate $R_\mathrm{B}$ and $M_\mathrm{B}$ from $T_\mathrm{eff,B}$ by using the empirical relations by \citet{Torres2010}.
These relations depend on $T_\mathrm{eff,B}$, $\log{g}_\mathrm{B}$ and $[Fe/H]_\mathrm{B}$.
We estimate a prior on \mbox{$\log{g}_\mathrm{B} \in \mathcal{N}(4.6,0.2)$} using the data by \citet{Pecaut2013} for our result \mbox{\TeffB{}~K}.
We further assume that \NstarA{} and B formed in the same system, and hence show similar metallicity. We hence set a metallicity prior of \mbox{$[Fe/H]_\mathrm{B} \in \mathcal{N}(0.,0.5)$}.

We find that \mbox{\RB{}~\Rsun{}} and \mbox{\MB{}~\Msun{}}.
Tab.~\ref{tab:MCMC_results} summarises all inferred results.
Fig.~\ref{fig:corner_derived} shows the inferred distributions for all parameters.

\subsection{Identifying \Nplanet{}}
\label{ss:Identifying Nplanet}

We use the MCMC chains and our inference of the systems dilution to calculate the properties of \Nplanet{}, the object orbiting \NstarA{}.
We can estimate the radius of \Nplanet{} directly from the MCMC samples of the ratio of radii, $R_C/R_A$, and the prior on $R_A$. We find \RC{}.
We estimate the mass of \Nplanet{} with the binary mass function $f$ for spectroscopic single-lined binaries:
\begin{equation}
f := \frac{P K^3 (1-e^2)^\frac{3}{2}}{2 \pi G} = \frac{M_\mathrm{C}^3 \sin{i}^3}{(M_\mathrm{C}+M_\mathrm{A})^2}
\end{equation}
We solve this equation for all MCMC samples ($P, K, i$) and the prior on $M_\mathrm{A}$. We find \MC{}.
Tab.~\ref{tab:MCMC_results} summarises all derived results.
Fig.~\ref{fig:corner_derived} shows the inferred distributions for all parameters.

\subsection{Identifying the binary orbit}
\label{ss:Identifying the binary orbit}

We find a significant difference in systemic RV for \NstarA{} and B (Tab.~\ref{tab:MCMC_results}), but it is not straightforward to use this to constrain the orbital separation; the likelihood space for $\Delta$RV spans orders of magnitudes and depends on its orbital parameters, which remain unconstrained (see Section~\ref{ss:Model of the RV between NstarA and NstarB}).
However, we can use the centroid information to constrain the projected separation. With an estimate of the distance to the system, this can be translated into an orbital separation.

We perform an SED fit to the magnitudes reported in Tab.~\ref{tab:stellar_properties} following the method presented in \citet{gillen17}. For modelling of the two stars \NstarA{} and \NstarB{} we use two separate stellar model spectra from PHOENIX.
As priors, we use our results of the spectral analysis for \NstarA{} ($R_A, T_\mathrm{eff,A}, \log{g}_\mathrm{A}$; see Tab.~\ref{tab:stellar_properties}), and the inferred posterior likelihoods for \NstarB{} ($R_B, T_\mathrm{eff,B}$; see Tab.~\ref{tab:MCMC_results}).
The prior on the surface gravity is again chosen to be $\log{g}_\mathrm{B} \in \mathcal{N}(4.6,0.2)$ (see Section~\ref{ss:Identifying NstarB}).
We here fix $[Fe/H]_\mathrm{A,B} = 0$ to avoid interpolation over wide ranges of metallicity (the PHOENIX spectra are given in steps of 0.5 in metallicity).
We find a distance of $d=$\distancepc{}~pc to the binary system.

Using this result, we can translate the projected sky separation of \locxsky{}~arcsec and \locysky{}~arcsec (constrained by the centroid data in our global MCMC model; see Tab.~\ref{tab:MCMC_results}) into AU. This gives a lower limit on the orbital semi-major axis of the binary, which is \mbox{$a_\mathrm{binary}$\orbsepAU{}~AU}.
Using Kepler's third law, we can determine that the binary period is \mbox{ $P_\mathrm{binary}$\orbperyr{}~yr }.
At this orbital separation we do not expect to detect any transit-timing variations (TTVs). Indeed, there was no evidence for any TTVs in the data.
The resulting binary orbit agrees well with typical scenarios of a planet in a binary system, further supporting the evidence for NGTS-3Ab.

\begin{table*}
  \centering
  \scriptsize
  \caption{Parameters of the \Ntarget{} system. Values and error bars are the median and 16th / 84th percentile of the MCMC posterior likelihood distributions.}
    \begin{tabular}{lllll}
\multicolumn{4}{c}{\textit{Fitted parameters (astrophysical)}} \\ 
\hline 
$\Delta x$ & Relative CCD x position of the blend & $85_{-87}^{+72}$ & milli-pixel \\ 
$\Delta y$ & Relative CCD y position of the blend & $133_{-71}^{+47}$ & milli-pixel \\ 
$D_\mathrm{0,A,NGTS}$ & Dilution of star A in NGTS & $0.434_{-0.032}^{+0.030}$ &  \\ 
$D_\mathrm{0,A,SPEC. g'}$ & Dilution of star A in SPECULOOS g' band & $0.409_{-0.038}^{+0.035}$ &  \\ 
$D_\mathrm{0,A,SPEC. r'}$ & Dilution of star A in SPECULOOS r' band & $0.432_{-0.034}^{+0.031}$ &  \\ 
$D_\mathrm{0,A,SPEC. i'+z'}$ & Dilution of star A in SPECULOOS i'+z' band & $0.449\pm0.027$ &  \\ 
$D_\mathrm{0,A,HARPS}$ & Dilution of star A in HARPS & $0.424_{-0.051}^{+0.045}$ &  \\ 
$P$ & Period & $1.6753728\pm0.0000030$ & days \\ 
$T_0$ & Epoch (HJD-2450000) & $7620.16790\pm0.00095$ & days \\ 
$R_\mathrm{planet}/R_\mathrm{A}$ & Ratio of radii & $0.1638\pm0.0045$ &  \\ 
$(R_\mathrm{A}+R_\mathrm{planet})/a$ & Sum of radii over the semi-major axis of the planet's orbit & $0.1792_{-0.0011}^{+0.0012}$ &  \\ 
$\cos{i}$ & Cosine of the inclination & $0.0077_{-0.0054}^{+0.0085}$ &  \\ 
$RV_\mathrm{sys,A}$ & Systemic RV of \NstarA{} & $8.566\pm0.049$ & km/s \\ 
$RV_\mathrm{sys,B}$ & Systemic RV of \NstarB{} & $9.032_{-0.064}^{+0.085}$ & km/s \\ 
$K$ & RV semi-amplitude & $-0.404\pm0.035$ & km/s \\ 
\hline 
\multicolumn{4}{c}{\textit{Fitted parameters (other)}} \\ 
\hline 
$A_\mathrm{CCF}$ & Maximal amplitude of the CCF profile & $0.52147_{-0.00070}^{+0.00076}$ &  \\ 
FWHM$_\mathrm{CCF,A}$ & FWHM of the CCF profile of \NstarA{} & $7.436\pm0.082$ & km/s \\ 
FWHM$_\mathrm{CCF,B}$ & FWHM of the CCF profile of \NstarB{} & $6.857_{-0.090}^{+0.078}$ & km/s \\ 
$\sigma(F_\mathrm{NGTS})$ & Error of the flux in NGTS & $10.247\pm0.079$ & mmag \\ 
$\sigma(\xi_x)$ & Error of the centroid in x & $12.114\pm0.097$ & milli-pixel \\ 
$\sigma(\xi_x)$ & Error of the centroid in y & $11.926\pm0.095$ & milli-pixel \\ 
$\sigma(F_\mathrm{SPEC. Callisto, g'})$ & Error of the flux in SPEC. Callisto g' band & $2.846_{-0.093}^{+0.099}$ & mmag \\ 
$\sigma(F_\mathrm{SPEC. Callisto, r'})$ & Error of the flux in SPEC. Callisto r' band & $3.03_{-0.12}^{+0.13}$ & mmag \\ 
$\sigma(F_\mathrm{SPEC. Europa, r'})$ & Error of the flux in SPEC. Europa r' band & $2.597_{-0.082}^{+0.087}$ & mmag \\ 
$\sigma(F_\mathrm{SPEC. Europa, i'+z'})$ & Error of the flux in SPEC. Europa i'+z' band & $2.512_{-0.080}^{+0.085}$ & mmag \\ 
$\sigma(F_\mathrm{SPEC. Io, i'+z'})$ & Error of the flux in SPEC. Io i'+z' band & $2.517\pm0.084$ & mmag \\ 
$\sigma$(RV) & Error of the RV & $0.043_{-0.010}^{+0.017}$ & km/s \\ 
$\sigma$(BIS) & Error of the BIS & $0.0317_{-0.0097}^{+0.015}$ & km/s \\ 
$\sigma$(FWHM) & Error of the FWHM & $0.084_{-0.023}^{+0.037}$ & km/s \\ 
$\sigma$(Contrast) & Error of the Contrast & $1.61_{-0.41}^{+0.66}$ &  \\ 
$\sigma$(CCF) & Error of the CCF 1 & $0.00322_{-0.00018}^{+0.00019}$ &  \\ 
$\sigma$(CCF) & Error of the CCF 2 & $0.00611_{-0.00033}^{+0.00036}$ &  \\ 
$\sigma$(CCF) & Error of the CCF 3 & $0.00574_{-0.00031}^{+0.00035}$ &  \\ 
$\sigma$(CCF) & Error of the CCF 4 & $0.00397_{-0.00022}^{+0.00025}$ &  \\ 
$\sigma$(CCF) & Error of the CCF 5 & $0.00436_{-0.00024}^{+0.00026}$ &  \\ 
$\sigma$(CCF) & Error of the CCF 6 & $0.00484_{-0.00027}^{+0.00030}$ &  \\ 
$\sigma$(CCF) & Error of the CCF 7 & $0.00518_{-0.00028}^{+0.00030}$ &  \\   
\hline
\multicolumn{4}{c}{\textit{Derived parameters for \NstarB{}}} \\ 
\hline 
$T_\mathrm{eff,B}$ & Effective temperature of \NstarB{} & $5230_{-220}^{+190}$ & K \\ 
$R_\mathrm{B}$ & Radius of \NstarB{} & $0.77_{-0.16}^{+0.22}$ & R$_\odot$ \\ 
$M_\mathrm{B}$ & Mass of \NstarB{} & $0.88_{-0.12}^{+0.14}$ & M$_\odot$ \\ 
$\rho_\mathrm{B}$ & Density of \NstarB{} & $1.13_{-0.23}^{+0.29}$ & $\rho_\odot$ \\ 
\hline
\multicolumn{4}{c}{\textit{Derived parameters for \Nplanet{}}} \\ 
\hline 
$R_\mathrm{planet}$ & Radius of the planet & $1.48\pm0.37$ & R$_\mathrm{J}$ \\ 
$M_\mathrm{planet}$ & Mass of the planet & $2.38\pm0.26$ & M$_\mathrm{J}$ \\ 
$\rho_\mathrm{planet}$ & Density of the planet & $0.31_{-0.15}^{+0.41}$ & $\rho_\mathrm{J}$ \\ 
$i$ & Inclination & $89.56_{-0.48}^{+0.31}$ & deg \\ 
$R_\mathrm{planet}/a$ & Planet radius over semi-major axis of the planet's orbit & $0.02523\pm0.00071$ &  \\ 
$R_\mathrm{A}/a$ & Radius of \NstarA{} over semi-major axis of the planet's orbit & $0.15398_{-0.00069}^{+0.00082}$ &  \\ 
$a$ & Semi-major axis of the planet's orbit & $5.0_{-1.0}^{+1.4}$ & R$_\odot$ \\ 
$T_\mathrm{1-4}$ & Total duration of transit & $138.15\pm0.82$ & min \\ 
$T_\mathrm{2-3}$ & Transit width & $98.82\pm0.63$ & min \\ 
$\delta_{\mathrm{undil.}} = (R_\mathrm{planet}/R_\mathrm{A})^2$ & Undiluted (real) depth of the transit & $2.68\pm0.15$ & per-cent \\ 
$b_\mathrm{tra}$ & Impact parameter of the transit & $0.050_{-0.035}^{+0.055}$ &  \\
\hline
\multicolumn{4}{c}{\textit{Derived parameters for the \Ntarget{} binary system}} \\ 
\hline 
$\Delta x_\mathrm{sky}$ & Relative sky position of the blend in x & $0.42_{-0.43}^{+0.36}$ & arcsec \\ 
$\Delta y_\mathrm{sky}$ & Relative sky position of the blend in y & $0.66_{-0.35}^{+0.23}$ & arcsec \\ 
$d$ & Distance to the system & \distancepc{} & pc \\
$a_\mathrm{binary}$ & Orbital separation between the stars & \orbsepAU{} & AU \\
$P_\mathrm{binary}$ & Orbital period of the binary stars & \orbperyr{} & yr \\
\hline
    \end{tabular}%
  \label{tab:MCMC_results}%
\end{table*}%
\vspace{2cm}


\section{Discussion}
\label{s:Discussion}

\subsection{\Ntarget{} as a cautionary tale of careful vetting}
\label{ss:Ntarget as a cautionary tail of careful vetting}
Only careful modelling of multi-colour photometry, centroids and RV CCF profiles and their bisectors enabled the verification of \Nplanet{}.
From single-colour photometry, centroids and RV measurements alone, \Nplanet{} would have been misclassified as an undiluted hot Jupiter orbiting an isolated G-type star.

On the other hand, a simpler consideration of the bisector correlation would have led to it being rejected as a planet. 
This finding is important to consider, as the bisector correlation is a common planet vetting criteria. 
It might have previously led to the erroneous rejection of bona-fide planets in unresolved binary systems.

We particularly raise caution that single-colour photometry alone, even if combined with precision centroiding, was not sufficient to identify the three-body nature of this system. 
Only if combined with multi-colour information and an analysis of the RV CCF profiles and BIS measurements we were able to unmask the hidden nature of this system.

We caution that scenarios like \Ntarget{} might be more common than currently anticipated.
Unresolved companions dilute exoplanet transit signals, biasing measured planetary quantities and potentially leading to miss-classification.
Diluted gas giant planets or Brown Dwarf companions in unresolved binary systems can also mimic Neptune-sized and rocky exoplanets.

\Ntarget{} is not resolved in \Gaia{} DR2, which was released during revision of this publication and is complete to an angular resolution of $0.4$\arcsec{}-$0.5$\arcsec{} separation \citep{Gaia2018}.
The non-identification of the companion in \Gaia{} DR2 is in agreement with the results of our global MCMC model, predicting a separation around the completeness limit of \Gaia{} DR2 (see Tab.~\ref{tab:MCMC_results}).
This highlights that hidden companion stars to exoplanet hosts in multi-star systems can remain unresolved in \Gaia{} DR2. 
Moreover, there was no sign of the companion in the SPECULOOS images, nor the HARPS guider images. 
It is hence crucial for transit surveys like \NGTS{} and the upcoming \TESS{} mission to account for the resolution limits of follow-up instruments and catalogues like \Gaia{} DR2.

The most robust way to identify hidden systems is a systematic lucky imaging or adaptive optics follow-up. Ideally, this would be conducted for any exoplanet system. 
In the case of \Ntarget{}, this will also allow to verify the accuracy of our modelling. 
We therefore aim to propose for high-resolution imaging of \Ntarget{}. 
Exploring this system further will place constraints on its binary companions, consequently refining the planetary parameters.

\subsection{Caveats and prospects}
\label{ss:Caveats and prospects}

\subsubsection{Priors on star A and B}
\label{sss:Priors on star A and B}

We draw our priors on star A from the HARPS spectral analysis. We caution that this is only correct if the flux from star A dominates the spectrum. In the case of similar luminosity of star A and B, the spectrum will be significantly influenced by both stars. The spectral analysis then approximately reflects a mean value between the two stars.
As our findings indicate that star B contributes to the overall spectrum, we might underestimate the effective temperature of star A.

Due to lack of any knowledge of star B, we have to assume it is a slow-rotating main-sequence star, which has the same prior on it's metallicity as star A. While reasonable, this assumption might cause a slight bias.

\subsubsection{Calibration of the HARPS CCF G2 mask}
\label{sss:Calibration of the HARPS CCF G2 mask}

There is no calibration of the HAPRS CCF G2 mask covering the entire range of effective temperatures from $3000-6000$~K. In particular, the model will profit from the following two calibrations:
\begin{align}
\mathrm{Contrast} = f \left( T_\mathrm{eff}, \log g, [Fe/H] \right),\\
\mathrm{FWHM} = f \left( T_\mathrm{eff}, \log g, [Fe/H] \right).
\end{align}
In Section~\ref{sss:Modelling the CCFs of blended systems} we studied these relations. While the current HARPS calibrations allow to constrain the relationship for the FWHM for effective temperatures $\gtrsim 3900$~K, there is no such calibration for the contrast. Our analysis of data from \citet{Sousa2008} only allowed to constrain the contrast for effective temperatures $\gtrsim 5000$~K. To avoid introducing a bias into the fit due to the break at this temperature, we decided to use uniform priors instead (which, however, by itself introduces some bias).

\section{Conclusion}
\label{s:Conclusion}

We report the disentanglement of a previously unresolved three-body system, \Ntarget{}, from multi-colour photometry, centroiding and radial velocity cross-correlation profiles. We highlight the discovery of \Nplanet{}, a potentially inflated hot Jupiter (\RC{}~\rjup{} and \MC{}~\mjup{}) in a \Nperiod{}~days orbit around the primary of an unresolved binary system. This provides an interesting testbed for planet formation, migration and orbital stability, as well as stellar multiplicity and metallicity.

Binary and triple systems are numerous. They frequently mimic exoplanet signals in photometric and radial velocity (RV) observations. 
We develop a thorough analysis framework, packaged in our {\scshape blendfitter} tool, to unmask such false positives and identify the true cause of detected signals. 
In particular, we analyse the photometric flux centroid as well as the RV cross-correlation functions and their bisectors.

\section*{Acknowledgements}
This research is based on data collected under the \NGTS{} project at the ESO La Silla Paranal Observatory. \NGTS{} is operated with support from the UK Science and Technology Facilities Council (STFC; project reference ST/M001962/1). Construction of the \NGTS{} facility was funded by the University of Warwick, the University of Leicester, Queen's University Belfast, the University of Geneva, the Deutsches Zentrum f\" ur Luft- und Raumfahrt e.V. (DLR; under the `Gro\ss investition GI-NGTS'), the University of Cambridge and STFC.
The research leading to these results has received funding from the European Research Council under the FP/2007-2013 ERC Grant Agreement number 336480 (SPECULOOS) and number 320964 (WDTracer), and from the ARC grant for Concerted Research Actions, financed by the Wallonia-Brussels Federation. This work was also partially supported by a grant from the Simons Foundation (PI Queloz, grant number 327127). 
This work has further made use of data from the European Space Agency (ESA) mission {\it Gaia} (\url{https://www.cosmos.esa.int/gaia}), processed by the {\it Gaia} Data Processing and Analysis Consortium (DPAC, \url{https://www.cosmos.esa.int/web/gaia/dpac/consortium}). Funding for the DPAC has been provided by national institutions, in particular the institutions participating in the {\it Gaia} Multilateral Agreement. 
Moreover, this publication makes use of data products from the Two Micron All Sky Survey, which is a joint project of the University of Massachusetts and the Infrared Processing and Analysis Center/California Institute of Technology, funded by the National Aeronautics and Space Administration and the National Science Foundation.
We make use of Python programming language \citep{Rossum1995} 
and the open-source Python packages
{\scshape numpy} \citep{vanderWalt2011}, 
{\scshape scipy} \citep{Jones2001}, 
{\scshape matplotlib} \citep{Hunter2007}, 
{\scshape pandas} \citep{McKinney2010}, 
{\scshape emcee} \citep{Foreman-Mackey2013}, 
{\scshape george} \citep{Ambikasaran2014}, 
{\scshape corner} \citep{Foreman-Mackey2016}, 
{\scshape seaborn} (\url{https://seaborn.pydata.org/index.html}), 
{\scshape pyastronomy} (\url{https://github.com/sczesla/PyAstronomy}), 
{\scshape pysynphot} \citep{Lim2013}, 
{\scshape limb-darkening} \citep{Espinoza2015}, 
and {\scshape eb} \citep{Irwin2011}.
The latter is based on the previous 
{\scshape JKTEBOP} \citep{Southworth2004a, Southworth2004b} and {\scshape EBOP} codes \citep{Popper1981}, 
and models by \cite{Etzel1981}, \cite{Mandel2002}, \mbox{\cite{Binnendijk1974a,Binnendijk1974b}} and \cite{Milne1926}.
We also make use of \textsc{IRAF}. \textsc{IRAF} is distributed by the National Optical Astronomy Observatory, which is operated by the Association of Universities for Research in Astronomy, Inc., under cooperative agreement with the National Science Foundation.
MNG is supported by the UK Science and Technology Facilities Council (STFC) award reference 1490409 as well as the Isaac Newton Studentship. 
LD acknowledges support from the Gruber Foundation Fellowship.
DJA, TL, DP, RGW, and PJW are supported by an STFC consolidated grant (ST/P000495/1). 
MG is FNRS-F.R.S. Research Associate. 
EJ is a senior research scientist at the Belgian FNRS.
MRG, MB are supported by an STFC consolidated grant (ST/N000757/1).

\bibliographystyle{mnras}
\bibliography{Guenther2018a_References}

\section*{Affiliations}
{\it
$^{c}$Astrophysics Group, Cavendish Laboratory, J.J. Thomson Avenue, Cambridge CB3 0HE, UK\\
$^{g}$Observatoire de Gen{\`e}ve, Universit{\'e} de Gen{\`e}ve, 51 Ch. des Maillettes, 1290 Sauverny, Switzerland\\
$^{w}$Dept.\ of Physics, University of Warwick, Gibbet Hill Road, Coventry CV4 7AL, UK\\
$^{ce}$Centre for Exoplanets and Habitability, University of Warwick, Gibbet Hill Road, Coventry CV4 7AL, UK\\
$^{k}$Astrophysics Group, Lennard-Jones Laboratories, Keele University, Staffordshire ST5 5BG, UK\\
$^{ks}$Space and Astronomy Department, Faculty of Science, King Abdulaziz University, 21589 Jeddah, Saudi Arabia\\
$^{ka}$King Abdullah Centre for Crescent Observations and Astronomy, Makkah Clock, Mecca 24231, Saudi Arabia\\
$^{li}$Space sciences, Technologies and Astrophysics Research (STAR) Institute, Universit\'e de Li\`ege, All\'ee du 6 Ao\^ut 17, Bat. B5C, 4000 Li\`ege, Belgium\\
$^{l}$Department of Physics and Astronomy, Leicester Institute of Space and Earth Observation, University of Leicester, LE1 7RH, UK\\
$^{d}$Institute of Planetary Research, German Aerospace Center, Rutherfordstrasse 2, 12489 Berlin, Germany\\
$^{tu}$Center for Astronomy and Astrophysics, TU Berlin, Hardenbergstr. 36, D-10623 Berlin, Germany\\
$^{fu}$Institute of Geological Sciences, FU Berlin, Malteserstr. 74-100, D-12249 Berlin, Germany\\
$^{q}$Astrophysics Research Centre, School of Mathematics and Physics, Queen's University Belfast, BT7 1NN Belfast, UK\\
$^{uc}$Departamento de Astronom\'ia, Universidad de Chile, Casilla 36-D, Santiago, Chile\\
$^{ci}$Centro de Astrof\'isica y Tecnolog\'ias Afines (CATA), Casilla 36-D, Santiago, Chile\\
$^{a}$Instituto de Astronom\'ia, Angamos 0610, D-1272709,Antofagasta, Chile
}

\section*{Appendix}
\renewcommand{\thefigure}{A\arabic{figure}}
\renewcommand{\thetable}{A\arabic{table}}
\renewcommand{\thesubsection}{\Alph{subsection}}

\begin{figure*}
\centering
 \includegraphics[width=2\columnwidth]{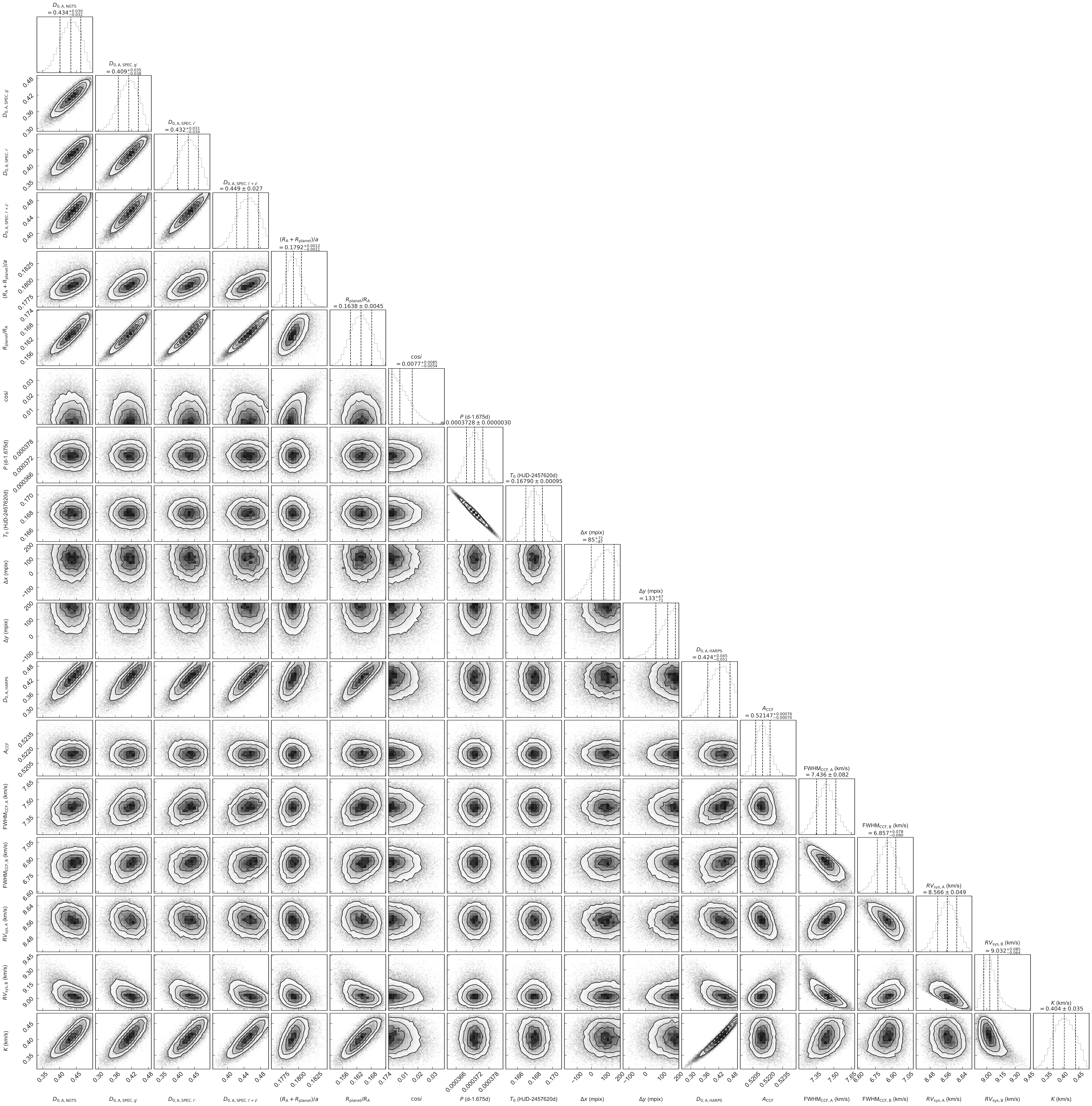}
 \caption{Posterior likelihood distributions for all astrophysical parameters of the MCMC fit to \Ntarget{}. For better visibility, the error scaling parameters are not shown here. Parameters are described in Tab.~\ref{tab:MCMC_results}.}
 \label{fig:corner}
\end{figure*}

\newpage
\begin{figure*}
\centering
 \includegraphics[width=2\columnwidth]{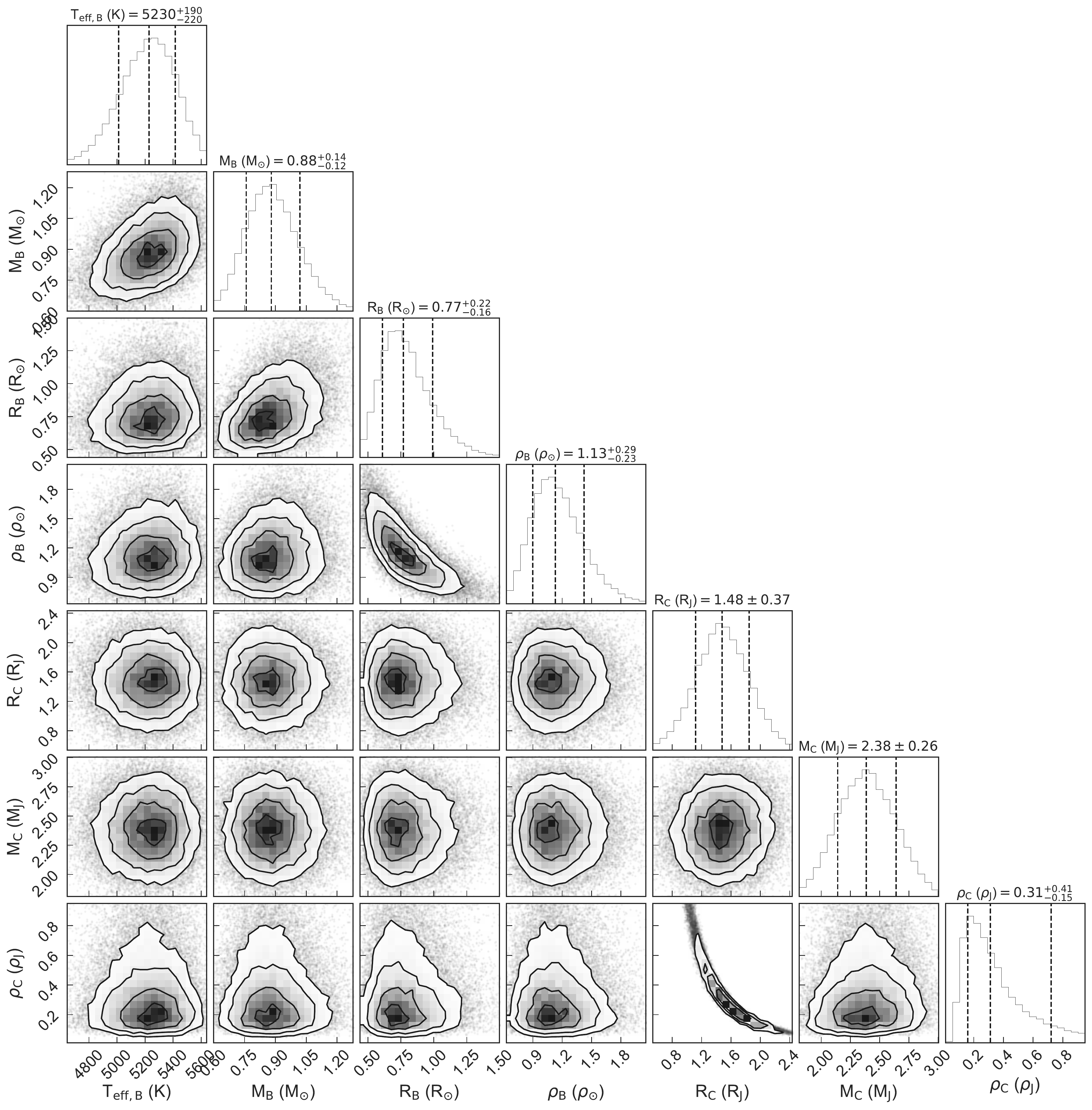}
 \caption{Likelihood distributions for the derived parameters for \NstarB{} and \Nplanet{}, as inferred from the results of our MCMC fit. Parameters are described in Tab.~\ref{tab:MCMC_results}.
 }
 \label{fig:corner_derived}
\end{figure*}

\bsp	
\label{lastpage}
\end{document}